\documentclass[twocolumn]{aastex63}

\shorttitle{}
\shortauthors{}
\graphicspath{{./}{figures/}}

\usepackage{amstext}
\usepackage{amsmath}
\usepackage{amssymb}
\usepackage{graphicx}

\newcommand{\beq}{\begin{equation}}
\newcommand{\eeq}{\end{equation}}
\newcommand{\beqar}{\begin{align}}
\newcommand{\eeqar}{\end{align}}

\newcommand{\Ms}{M_\ast}
\newcommand{\Rs}{R_\ast}
\newcommand{\Ma}{M_{\rm a}}

\newcommand{\Tc}{T_{\rm c}}
\newcommand{\Xd}{X_{\rm d}}
\newcommand{\kd}{\kappa_{\rm d}}
\newcommand{\vw}{v_{\rm w}}

\newcommand{\gloss}{\gamma_{\rm loss}}

\begin{document}

\title{Dusty, Self-Obscured Transients from Stellar Coalescence}
\correspondingauthor{Morgan MacLeod}
\email{morgan.macleod@cfa.harvard.edu}

\author[0000-0002-1417-8024]{Morgan MacLeod}
\affiliation{Center for Astrophysics $\vert$ Harvard $\&$ Smithsonian 
60 Garden Street, MS-16, Cambridge, MA 02138, USA}

\author[0000-0002-8989-0542]{Kishalay De}
\altaffiliation{NASA Einstein Fellow}
\affiliation{MIT-Kavli Institute for Astrophysics and Space Research, 77 Massachusetts Avenue, Cambridge, MA 02139, USA}

\author[0000-0003-4330-287X]{Abraham Loeb}
\affiliation{Center for Astrophysics $\vert$ Harvard $\&$ Smithsonian 
60 Garden Street, MS-16, Cambridge, MA 02138, USA}

\begin{abstract}
We discuss the central role that dust condensation plays in shaping the observational appearance of outflows from coalescing binary systems. As binaries begin to coalesce, they shock-heat and expel material into their surroundings. Depending on the properties of the merging system, this material can expand to the point where molecules and dust form, dramatically increasing the gas opacity. We use the existing population of Luminous Red Novae (LRNe) to constrain the thermodynamics of these ejecta, then apply our findings to the progressive obscuration of merging systems in the lead in to their coalescence. Compact progenitor stars near the main sequence or in the Hertzsprung gap along with massive progenitor stars have sufficiently hot circumstellar material to remain unobscured by dust. By contrast, more extended, low-mass giants should become completely optically obscured by dust formation in the circumbinary environment.  We predict that 30--50\% of stellar coalescence transients for solar-mass stars will be dusty, infrared-luminous sources. Of these, the optical transients may selectively trace complete merger outcomes while the infrared transients trace common envelope ejection outcomes. 
\end{abstract}


\section{Introduction}

 Common envelope phases, with their possible outcomes of stellar mergers or envelope ejection  are transformational episodes in multiple star lifetimes. These phases modify the mass, luminosity, and spectral distributions of stellar populations \citep{2017PASA...34....1D}. They are crucial in shaping the evolutionary history of many massive stars, which are especially likely to have close companions \citep{2012Sci...337..444S,2013ARA&A..51..269D,2014ApJ...782....7D}. And finally, stellar mergers and common envelope ejections are thought to be crucial in producing exotic stellar outcomes, like rapidly rotating and highly magnetized stars \citep[e.g.][]{2007MNRAS.375..909S,2013ApJ...764..166D,2016MNRAS.457.2355S}, or compact binaries that go on to merge under the influence of gravitational radiation \citep[e.g.][]{2012ApJ...759...52D,2013A&ARv..21...59I,2014LRR....17....3P,2020PASA...37...38V,2021A&A...650A.107M}. 

The growing number of transient surveys across the electromagnetic and gravitational wave spectra are offering a new channel for insight into these events. Much of traditional astronomical inference about these objects has come from comparing populations of binary and multiple stars before and after presumed interactions \citep[e.g.][]{1976IAUS...73...75P,1976IAUS...73...35V}. With time-domain surveys, the possibility of observing these events as galactic or extragalactic transients has become feasible. Catching these events in action is poised to allow us to make direct connections between stellar binaries (and multiples), the transients they produce as they coalesce, and the outcomes of their interactions. 

This paper attempts a step toward those inferences on the basis of the population of Luminous Red Novae (LRNe). LRNe have been associated with stellar mergers, largely through comparison to the particularly clear case of V1309 Sco -- an eclipsing binary system with a decaying orbit that went into outburst and emerged as a single, thermally expanded star \citep{2010A&A...516A.108M,2011A&A...528A.114T,2014AJ....147...11M,2015A&A...580A..34K,2016A&A...592A.134T,2019MNRAS.486.1220F}.  There is now a small sample of LRNe ($\sim 10$ systems) with known progenitor properties along with detailed observations of the outbursts themselves  \citep[for a recent tabulation, see][and Section \ref{sec:lrn}]{2022arXiv220210478M}.

As the ejecta of LRNe expands, they cool, producing progressively redder emission.  As a natural consequence of this cooling, these ejecta eventually form molecules and dust \citep[see][for detailed analysis of the galactic sources V4332 Sgr, V1309 Sco, and V838 Mon]{2018A&A...617A.129K,2021A&A...655A..32K}. Dust formation episodes mark each of the LRNe, and these sources become highly infrared luminous in their immediate aftermath as dusty shells enshroud their ongoing evolution \citep[e.g.][]{2013MNRAS.431L..33N,2014AJ....147...11M,2014A&A...569L...3C,2016A&A...592A.134T,2020MNRAS.496.5503B,2020RNAAS...4..238M}. Further, the contribution of dust formed in common envelope ejecta is thought to be similar to that formed by single evolved stars \citep{2013ApJ...768..193L,2013ApJ...777...23Z,2015RAA....15...55W}. While dust formation appears to be ubiquitous, the shocks that heat LRNe ejecta depend on the particular system's properties. The characteristic condensation temperature ($\sim 10^3$~K) for dust grains largely does not. We therefore expect differences to emerge in the efficiency of dust formation and the degree of optical obscuration that results. 

In this paper, we draw on scalings from the observed LRNe to constrain the uncertain thermodynamics of LRNe ejecta (Section \ref{sec:lrn}). We model the phases of pre-outburst mass loss into the circumbinary environment \citep{2014ApJ...788...22P,2016MNRAS.455.4351P,2016MNRAS.461.2527P,2017ApJ...850...59P,2018ApJ...863....5M,2018ApJ...868..136M,2020ApJ...893..106M,2020ApJ...895...29M}, that are believed to precede outburst from stellar coalescence (Section \ref{sec:model}). Our models predict the circumbinary mass and temperature distribution, and allow us to convert this to order-of-magnitude estimates for the degree of dust obscuration, which varies widely depending on the merging system's properties (Section \ref{sec:results}). The qualitative picture that emerges is one of a bifurcation of transients between the optical and infrared, depending on the properties of progenitor. In Section \ref{sec:conclusions}, we conclude. 

\section{Dust Formation in Merger Ejecta}\label{sec:lrn}

First, we draw on constraints from the subset of the population of LRNe with progenitor detections to understand the  cooling of ejecta and dust condensation in these events. 

\subsection{Luminous Red Nova Sample}\label{sec:sample}
A growing number of progenitor systems of LRNe have been discovered. This allows us to compare the binary systems (or at least the primary stars in these binary systems) to the transients that they produce. We make use of the modeled progenitor masses, radii, luminosities, and effective temperatures reported in the literature for eight transients: V4332 Sgr \citep{1999AJ....118.1034M}, V838 Mon \citep{2002A&A...389L..51M,2003Natur.422..405B,2005A&A...436.1009T,2005A&A...441.1099T}, V1309 Sco \citep{2010A&A...516A.108M,2011A&A...528A.114T}, NGC4490-OT2011 \citep{2016MNRAS.458..950S}, M31 LRN 2015 \citep{2015ApJ...805L..18W,2015A&A...578L..10K,2017ApJ...835..282M,2020MNRAS.496.5503B}, M101 LRN 2015 \citep{2017ApJ...834..107B}, SN Hunt 248 \citep{2015MNRAS.447.1922M}, and AT2018 bwo \citep{2021A&A...653A.134B}. The extragalactic transients AT 2018hso \citep{2019A&A...632L...6C}, AT 2019zhd  \citep{2021A&A...646A.119P}, AT 2020hat and AT 2020kog \citep{2021A&A...647A..93P} also have pre-outburst absolute magnitudes and colors, but no comparisons to stellar model tracks have been made. 

The galactic transients OGLE-2002-BLG360 \citep{2013A&A...555A..16T} and CK Vul (1670) \citep{1985ApJ...294..271S}, are both possibly associated with a stellar merger origin despite having light curves that are somewhat distinct from the other LRNe \citep[e.g. as discussed by][]{2013A&A...555A..16T}. Unfortunately, the physical properties of the progenitors of these outbursts remain uncertain. OGLE-2002-BLG360 has a number of interesting properties in the context of dusty transients, which we discuss in Section \ref{sec:blg360}.

It is important to highlight that each of these progenitor identifications suffers from the uncertainty of model dependence. In particular, mass transfer could potentially induce significant departures in stellar appearance relative to an isolated-star model \citep{2021A&A...653A.134B}. For our analysis, we adopt the tabulated properties of systems in \citet{2022arXiv220210478M}.

\subsection{Molecule and Dust Condensation in LRNe}

The light curves of LRNe exhibit a progressive reddening after peak luminosity, which traces the cooling emission of their ejecta. This evolution was first traced in exquisite detail during in the 2002 outburst of V838 Mon \citep{2002A&A...389L..51M,2002MNRAS.336L..43K,2003Natur.422..405B,2005A&A...436.1009T}. The source was quickly attributed to a possible stellar coalescence origin \citep{2003ApJ...582L.105S,2006A&A...451..223T,2006MNRAS.373..733S}.  Early spectroscopy of V838 Mon showed H$\alpha$ emission that was soon accompanied by broadened P-Cygni absorption lines tracing the ejecta velocity \citep{2002A&A...389L..51M}.  Three years after the outburst, high-resolution spectra \citet{2009ApJS..182...33K} show molecular absorption outside the remnant's photosphere, still broadened by outflow into P-Cygni profiles.  Early near-infrared spectra traced the emergence of CO and other molecules \citep{2002A&A...395..161B}, and an infrared excess in the object's spectral energy distribution was also noted \citep[e.g.][]{2005A&A...436.1009T}. Molecular CO emission in the sub-millimeter bands has since been a valuable tracer of the ejecta \citep{2007A&A...475..569K}, spatially resolved approximately 5 years after the outburst \citep{2008A&A...482..803K}. It has now been observed with Atacama Large Millimeter Array (ALMA), where the interaction of the ejecta with a tertiary component in the system is seen in continuum dust emission and a variety of carbon and sulfur oxide lines \citep{2021A&A...655A..32K}. Further recent constraints on the long-term composition and properties of the dusty ejecta have been derived from Spitzer space telescope and SOFIA observations \citep{2021AJ....162..183W}. 

Subsequent transients have also been observed to undergo epochs of significant dust formation, simultaneous with the dimming of the optical light curve. For example, \citet{2013MNRAS.431L..33N} and \citet{2016A&A...592A.134T} showed the significance of an evolving spectral energy distribution in V1309 Sco, with a significant shift toward infrared dominance as the optical light faded  \citep[see, for example, Figure 7 of][]{2016A&A...592A.134T}. Meanwhile V1309 Sco's optical and sub-millimeter spectra traced similar evolution of molecular formation and absorption seen in V838 Mon and V4332 Sgr, strengthening the connection between these transients \citep{2015A&A...580A..34K,2018A&A...617A.129K}.  V1309 Sco shows evidence for warm dust ($\sim 10^3$~K) even prior to the outburst, indicating that dust formation was ongoing, but at a lower level than in the aftermath of the merger \citep{2016A&A...592A.134T}.  As a decade-older twin of V838 Mon and V1309 Sco, V4332 Sgr's remnant has proved valuable in understanding the temporal evolution of these sources \citep[e.g.][]{2003ApJ...598L..31B,2004ApJ...604L..57B,2005A&A...439..651T,2010A&A...522A..75K,2011A&A...527A..75K,2013A&A...558A..82K,2015A&A...578A..75T,2018A&A...617A.129K}.  The more recently-discovered extragalactic LRNe all share similar cooling and molecule and dust-forming properties, which are described in detail in the references in Section \ref{sec:sample} as central evidence for their association with a stellar-coalescence origin.

Theoretically, there have been some efforts to model dust formation in common envelope ejecta. Models of dust formation, usually adapted from the asymptotic giant branch (AGB) star context, have been applied to ejecta kinematics and thermodynamics thought to mimic those of merger and common envelope ejecta \citep{2013ApJ...768..193L,2013ApJ...777...23Z}.  These models estimate dust masses as a function of radius under different model parameters, and also predict the population contribution of common envelope ejecta to the total dust content of the interstellar medium \citep{2013ApJ...768..193L,2015RAA....15...55W}. \citet{2018MNRAS.478L..12G} have pointed out that the adiabatic cooling of common envelope ejecta makes dust formation inevitable, and that the increase in opacity associated with this transition may allow radiation pressure to play an important role in driving mass ejection, as in dust driven winds from pulsating AGB stars. \citet{2019MNRAS.489.3334I,2020MNRAS.497.3166I} perform a related, but more detailed analysis of the thermodynamic trajectories of simulated common envelope phases, and the implications for dust condensation in a dynamic model. As hinted by the V1309 Sco data \citep{2016A&A...592A.134T}, they find that warm dust condenses both in early, slow outflow from the phase of Roche lobe overflow preceding the merger, and in the later common envelope ejecta, with slightly different properties of grain size distribution \citep{2020MNRAS.497.3166I}. 

Dust formation appears to be a ubiquitous and inevitable consequence of the cooling of expanding  common envelope ejecta. The composition and quantity of this dust remain areas of active observational study, and are sensitive to different parameterizations of the uncertain ejecta thermodynamics \citep[e.g.][]{2013ApJ...768..193L}.  In the next section we focus on the condensation of warm dust at a characteristic temperature of $\sim 10^{3}$~K as a means of tracing the constraining the uncertain thermodynamics of LRNe ejecta.

\subsection{Constraints on Ejecta Thermodynamics}

The thermodynamics of cooling LRNe ejecta is uncertain. It is dictated by the equation of state and ionization state transitions of hydrogen and helium in the relevant temperature range, as well as by internal shock heating and radiative cooling. Thus, it is not clear that the ejecta temperatures should follow tracks of adiabatic cooling as they expand. By comparing the LRNe population to general power-law temperature profiles, $T(r) \propto r^{-\beta}$, we can constrain the normalization and slope of the temperature structure of these ejecta. 

The model power-law temperature as a function of radius is 
\beq
T(r) = T_0 \left(\frac{r}{R_0}\right)^{-\beta}
\eeq
where $R_0$ is the base of the outflow. 
We parameterize the temperature at the base of the outflow as
\beq
T_0 = f_{\rm vir} T_{\rm vir}
\eeq
where $f_{\rm vir} = T_0 /T_{\rm vir}$ is the fraction of the Virial temperature, $T_{\rm vir}$, of the primary star at radius $R_0$,
 \beq
 T_{\rm vir}(R_0) = \frac{\mu m_{\rm p}}{k_B} \frac{G M_0}{R_0},
 \eeq
 where we will adopt $M_0 \rightarrow \Ms$ and $R_0 \rightarrow \Rs$. 
Thus,
\beq
T_0 \sim 3\times10^4 \left( \frac{f_{\rm vir}}{0.02} \right)  \left( \frac{ M_0 }{M_\odot} \right) \left( \frac{R_0}{16R_\odot} \right)^{-1}~{\rm K},
\eeq  
normalizing to example parameter values. Here we note that $f_{\rm vir}$ is thought to be related to the temperature to which material is shock-heated before it is expelled, but it could equivalently represent the original temperature of unshocked ejecta. 

This temperature profile implies that material will condense to form dust grains at a radius of the order of 
\beq\label{rdust}
r_{\rm dust} = \left( \frac{T_{\rm dust}}{T_0} \right)^{-1/\beta} R_0,
\eeq
where $T_{\rm dust}\sim 10^3$~K is the temperature at which significant quantities of dust form. This transition induces a major opacity increase, that changes the spectral energy distribution of the LRN from optical to infrared dominated.

During the outburst phase of a stellar coalescence transient, the luminosity and temperature of of the coalescing binary increase dramatically. The source of this increased output is the dissipation of orbital energy into the surrounding gaseous envelope. One possible effect of increasing the central object's luminosity is modifying the thermodynamic properties of the circumbinary material. In thermodynamic equilibrium, the dust formation will occur at
\beq
r_{\rm dust,eq}= \left( \frac{L}{4\pi \sigma T_{\rm dust}^4} \right)^{1/2}.
\eeq 
Comparing $r_{\rm dust}$ as determined by the internal shock temperature to $r_{\rm dust,eq}$ determined by the equilibrium thermodynamics allows us to determine the relative importance of the increased outburst luminosity in inhibiting dust formation.

\begin{figure}[tbp]
\begin{center}
\includegraphics[width=0.49\textwidth]{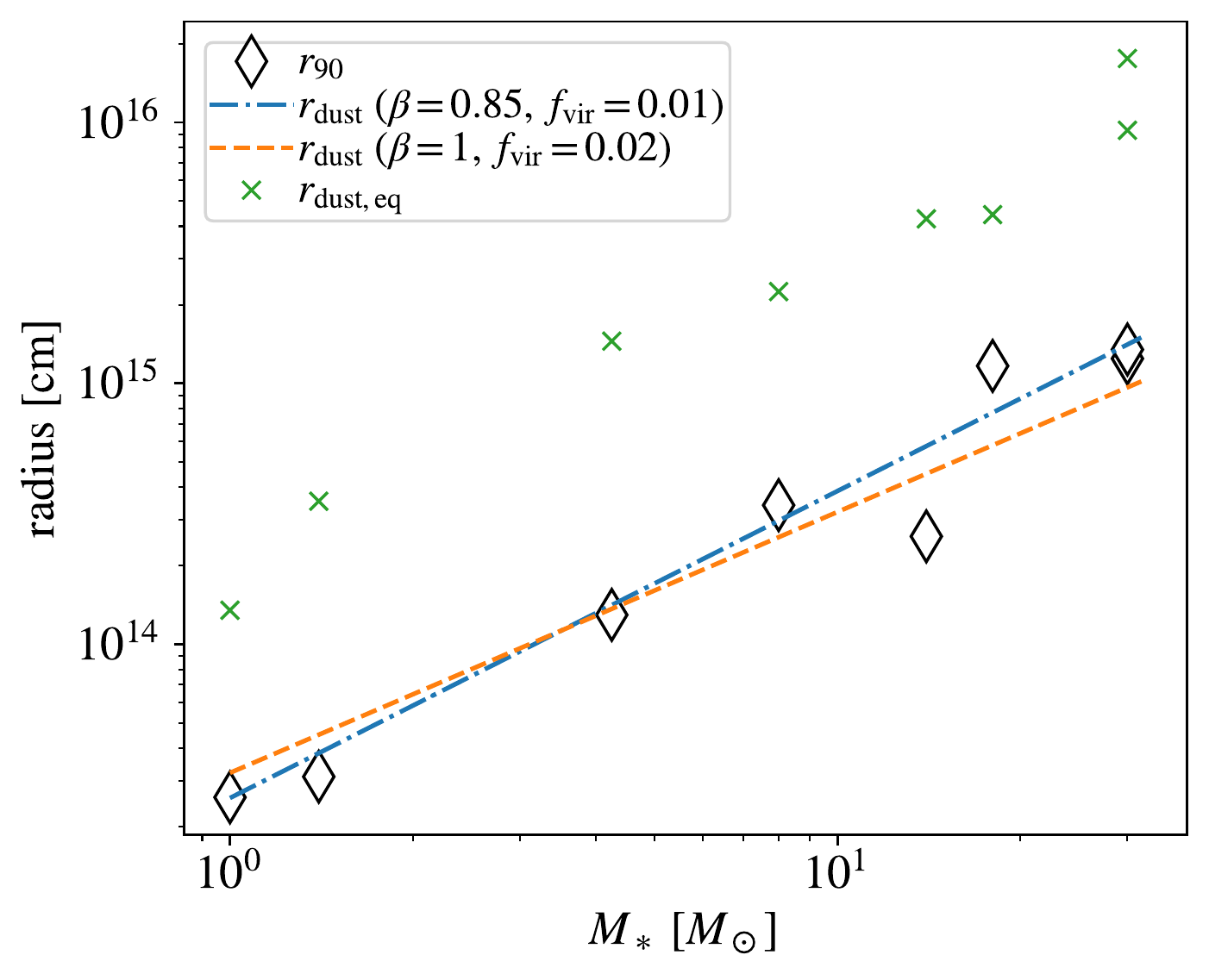}
\caption{The characteristic radius of dust formation in LRNe  as measured by $r_{90}$ and as predicted by various models. The thermal equilibrium dust formation radii are much larger than $r_{90}$ for all of the observed systems, which are better modeled by Virialized ejecta with $f_{\rm vir} \sim 10^{-2}$ and $\beta \sim 1$. The constant Virial fraction implied by $\beta \sim 1$ is highly suggestive of internal shocks shaping the ejecta thermodynamics. } 
\label{fig:r90}
\end{center}
\end{figure}

We apply these considerations to the observed LRNe in Figure \ref{fig:r90}. Here we compute 
\beq
r_{90} = t_{90} v_{\rm obs},
\eeq
for each transient, where $t_{90}$ is the timescale encompassing 90\% of the optical emission and $v_{\rm obs}$ is the velocity inferred from doppler-broadened emission lines \citep[$L_{90}$ is the mean luminosity over $t_{90}$,][]{2022arXiv220210478M}.   This represents the approximate radius that ejecta reach during the optical duration of the light curve. After this time, LRNe tend to form dust and become extremely infrared luminous \citep{2022arXiv220210478M}. We therefore suggest that, to order of magnitude, $r_{90}$ is the radial location where dust formation occurs, modifying the LRN spectral energy distribution.  We see that for each LRN transient, $r_{90} \ll r_{\rm dust,eq}$, adopting $L=L_{90}$. This indicates  that the thermal equilibrium temperature is not established in the ejecta, and does not inhibit dust formation at radii smaller than $r_{\rm dust,eq}$. Next, we compare models of $r_{\rm dust}$ based on the temperature resulting from internal shocks. We find that $\beta \sim 1$ and $f_{\rm vir}\sim 10^{-2}$ provide a good description of the data with $M_0=\Ms$ and $R_0= \Rs$. 

This indicates that ejecta are only weakly thermalized at the ejection radius, but that their temperature profile decays more gradually than the equation of state alone would indicate, most likely due to internal shock heating \citep[e.g.][]{2016MNRAS.455.4351P,2017MNRAS.471.3200M}.  
A temperature profile of $\beta =1$ corresponds to gas specific energy proportional to the gravitational potential, in which case the Virial fraction is constant with radius, as would be expected for an internal-shock origin for the temperature profile when the ejection velocity is similar to the local escape velocity.  

Compared to models of common envelope ejecta dust formation, this result is in tension with previous assumptions. \citet{2013ApJ...768..193L} assume shallower temperature profiles of $\beta = 0.2$--0.4, along with a shallower density profile $\rho \propto r^{-3/2}$, noting that different assumptions about the temperature profile in that range could induce seven orders of magnitude difference in the total modeled dust mass (this is likely, in part, due to the divergent density profile that is chosen -- if most of the dust is at large radii, the temperatures at those radii become very important). \citet{2019MNRAS.489.3334I} and \citet{2020MNRAS.497.3166I} adopt a very different kinematic and thermodynamic structure motivated by late-stage adiabatic expansion of unbound ejecta in their common envelope simulation. They find an expanding shell of ejecta decreases in density as $\rho \propto r^{-3}$, and at sufficiently late times, cools adiabatically, such that $T\propto r^{-2}$. However, their models also show evidence of the specific entropy increases at earlier times that trace shock heating. Taken together, these results could suggest  that internal shocks are important over some temporal and radial regime as the ejecta expand and cool toward the dust condensation temperature but that they asymptotically approach homologous, adiabatic expansion at larger radii.

\section{Pre-Outburst Mass Loss Model}\label{sec:model}

In this section, we describe a simplified model for the mass loss that precedes binary coalescence that is motivated by constraints from LRNe outbursts and by modeling of these phases.

\subsection{Coupled Mass Loss and Orbital Decay}

We define a model binary system consisting of an evolving primary star of mass $\Ms$ and radius $\Rs$, which is interacting with a more compact companion of mass $\Ma=q\Ms$, where $q$ is the binary mass ratio. To model the runaway orbital decay associated with dynamically unstable mass transfer, we use the {\tt RLOF} python package \citep{RLOF1.1}. {\tt RLOF} adopts numerical coefficients based on hydrodynamic simulations of runaway binary coalescence \citep{2020ApJ...893..106M,2020ApJ...895...29M} and solves the ordinary differential equations of binary orbital evolution in the point-mass binary limit. 

We initialize our model system when the primary star is slightly overflowing its Roche lobe,  $a=0.999a_{\rm RL}$, where $a$ is the orbital semi-major axis and $a_{\rm RL}$  is the semi-major axis of Roche lobe overflow as estimated by the \citet{1983ApJ...268..368E} approximation.  We integrate the coupled equations of non-conservative mass loss from a binary system and orbital angular momentum conservation, which reduce to 
\begin{align}
\dot a &=  -2 a \frac{\dot \Ms}{\Ms} \left[ 1 - \left( \gloss + {1\over 2}\right) {\Ms \over M}  \right],  \label{adot} \\
\dot \Ms & = -  \alpha   {\Ms \over P_{\rm orb} } \left( {\Rs - R_L \over \Rs} \right)^{n+{3\over 2}}, \label{mdot}
\end{align}
where $M=\Ms + \Ma$, and 
\beq
\gloss = {l_{\rm loss}  \over l_{\rm bin}},
\eeq
is a dimensionless specific angular momentum of material leaving the binary. It is the ratio of the specific angular momentum, $l_{\rm loss}$, of lost material to the specific angular momentum of the binary, $l_{\rm bin} = \Ms \Ma / M^2 \sqrt{GMa}$. In the expression for $\dot \Ms$, $\alpha$ is a coefficient of order unity, the orbital period is 
\beq
P_{\rm orb} = 2\pi \left( \frac{a^3}{GM} \right)^{1/2},
\eeq
$R_L$ is the the Roche lobe radius \citep{1983ApJ...268..368E}, and $n=1+1/\Gamma$ is the polytropic index of the primary star. Equation \eqref{mdot} is from  \citet{1972AcA....22...73P}, and has been numerically verified in hydrodynamic simulations by \citep{2018ApJ...863....5M,2020ApJ...893..106M}, and represents adiabatic mass loss from a polytropic donor star. We caution that in cases where the thermal adjustment of the mass-shedding layers of the donor star are important, the approximation of equation \eqref{mdot} will be inaccurate. 

In what follows, to provide context for the qualitative trends across many binaries, we make some uniform assumptions for all of the binary systems we model. We assume all primary stars have $n=3/2$ polytropic index that enters into equation \eqref{mdot}, and we further assume that the adiabatic response of the donor star is to maintain constant radius upon mass loss.  In reality, the structure of donor stars depends on their evolutionary state, with most objects ranging between effective polytropic indices of $4/3\leq n \leq 5/3$; the choice of $n=3/2$ represents an intermediate value. The response of systems to mass loss is more complex because, in addition to the adiabatic response of a donor star \citep{1987ApJ...318..794H}, thermal adjust modifies the mass--radius relation dependent upon the mass transfer rate itself. More tailored parameters can be easily specified with {\tt RLOF}, or similar models that account for the thermal and nuclear evolution of the primary star could be completed with the binary module of MESA \citep{2011ApJS..192....3P,2013ApJS..208....4P,2015ApJS..220...15P,2018ApJS..234...34P,2019ApJS..243...10P}, for example as recently computed in a similar context by \citet{2021A&A...653A.134B}. 

\subsection{Circumbinary Material}
The kinematics and thermodynamics of material in the circumbinary environment are complex and multidimensional. The continuous orbital motion launches spiral shocks through outflowing material, which provide heat and redistribute angular momentum. Early, slower mass loss trails away from the binary concentrated toward the system's orbital plane and may be either bound \citep{2016MNRAS.461.2527P,2018ApJ...868..136M,2020ApJ...895...29M}, or weakly unbound \citep{2016MNRAS.455.4351P,2017ApJ...850...59P,2019MNRAS.489..891H}. Later, faster ejecta encounter this torus, and are reshaped into bipolar outflows \citep{2018ApJ...868..136M}.

\subsubsection{Torus Model}\label{sec:torus}

If the pre-coalescence mass loss is bound to the binary system, as found in the hydrodynamic simulations of \citet{2018ApJ...868..136M,2020ApJ...895...29M}, it forms a thick torus surrounding the binary. Within this torus, gas is shock heated by the continual spiral outflow from the binary. \citet{2020ApJ...895...29M} found that these shocks also redistribute angular momentum, such that the torus has nearly constant specific angular momentum determined by the total mass loss and total angular momentum change of the decaying orbit. An analytic solution for a hydrostatic toroidal configuration with constant specific angular momentum can be derived if we assume a polytropic equation of state $P=K\rho^\gamma$. Then,
\beq
\rho(R,z) = \left[ \frac{\gamma-1}{K\gamma} \left( \frac{GM}{r} - \frac{l^2}{2R^2} - \frac{GM}{R_0} + \frac{l^2}{2R_0^2} \right) \right]^{\frac{1}{\gamma-1}},
\eeq
where $R$ and $z$ are cylindrical coordinates relative to the central binary mass, $M$. In this parameterization, $l$ is the torus specific angular momentum and $R_{\rm t}$ is the outer radius of the torus (see appendix B of \citet{2020ApJ...895...29M} for the derivation of this expression). 

On the basis of hydrodynamic simulation results, \citet{2020ApJ...895...29M} provide fiducial torus parameters for this torus model as it is implemented in {\tt RLOF}. We set $l=\Delta L/\delta m$ where $\Delta L$ is the cumulative binary angular momentum change, and $\delta m$ is the mass change. This implies that all mass and angular momentum lost from the binary go to forming the torus. We set 
\beq
R_{\rm t}  = 200 q_0^{0.64}  R_*,
\eeq
 where $q_0$ is the initial binary mass ratio. This implies that the extent of the circumbinary material scales with the primary star radius.

Finally, the specification of the torus polytropic constant, $K$,  and index, $\gamma$, closes the model. These are linked to the temperature structure via an ideal gas equation of state $T=(P/\rho)(\mu m_{\rm p}/k_B)$, where we adopt $\mu \approx 1$.   We choose $\gamma=4/3$, then the polytropic constant $K$ is calculated for self-consistency between $R_{\rm t}$ and the integral torus mass. The choice of $\gamma=4/3$ is motivated by two factors. First, this is the best fit to   \citet{2020ApJ...895...29M}'s hydrodynamic simulation models, where internal shocks (rather than the gas' adiabatic index)  are the critical factor in establishing this entropy profile. Second, this implies $T(R,0) \propto R^{-1}$ in the bulk of the torus, just as noted in Section \ref{sec:lrn} for the Virialized ejecta from LRNe. We note that this temperature probably represents the  post-shock temperature in the wake of passing internal shocks, $T_{\rm sh}$, and can be modified by 
radiative cooling and diffusion in the intervals between the passage of shocks \citep[e.g.][]{2016MNRAS.455.4351P}.

The Virial fraction to which circumbinary material is shock-heated is proportional to the square of the ratio of the velocity dispersion of internal shocks over to the characteristic velocity at a given radius,  $f_{\rm vir}\sim \Delta v^2 / (G \Ms/r)$ \citep[e.g.][]{2016MNRAS.455.4351P,2020ApJ...895...29M}.  The normalization of our models shows that $f_{\rm vir}$ is a function of binary mass ratio, with $f_{\rm vir}\sim 0.25$ for $q=1/3$, and $f_{\rm vir}\sim 0.15$ for $q=0.01$ \citep{2020ApJ...895...29M}. We note that this is significantly higher $f_{\rm vir}$ than found for the outburst phase in Section \ref{sec:lrn}; this is likely because the slow, early outflow self-intersects much more dramatically as it decelerates in the gravitation potential than does the faster, more homologous late ejecta which are being modeled in Figure \ref{fig:r90}.

\subsubsection{Wind Model}\label{sec:wind}

If, rather than being bound to the binary, the circumbinary material is unbound and expands outward with an asymptotic velocity $\vw$, we can adopt a different model to describe the circumbinary  distribution. We imagine that the most likely scenario is a relatively slow, equatorially-concentrated outflow, in which internal shocks again play a central role in the thermodynamic structure \citep{2016MNRAS.455.4351P}.  

In this model, we specify the wind velocity to be a fraction of the stellar escape velocity,
\beq
v_w = f_v \left( 2G \Ms \over \Rs\right)^{1/2},
\eeq
where, motivated by \citet{2016MNRAS.455.4351P}, we choose $f_v = 0.25$ to represent a slow outflow solution. 
We use the time-evolving mass loss rate $\dot \Ms(t)$ from the {\tt RLOF} solution, along with the substitution 
\beq \label{rwind}
r= \Rs + \vw \Delta t, 
\eeq
where $\Delta t = t_{\rm merge}-t$ is the time before binary merger, in our case, when $a=\Rs$.
Each shell of the circumbinary distribution therefore corresponds to a certain time of emission from the binary. Then
\beq
\rho (r) \approx - \frac{\dot \Ms(r)}{4\pi r^2 \vw},
\eeq 
where $\dot \Ms(r)$ is related to $\dot \Ms(t)$ by equation \eqref{rwind} and the sign comes from the fact that $\dot \Ms < 0$.   We note that the time dependence of $\dot M(t)$ implies a density slope steeper than $r^{-2}$ surrounding the binary. Under our fiducial assumptions, we find scaling similar to $\rho \propto r^{-3.5}$. 

We use the Viral factor to set the post-shock temperature structure of the outflow,
 \beq
 T_{\rm sh}(r) = f_{\rm vir} T_{\rm vir} \left( \frac{r}{\Rs} \right)^{-1}
 \eeq 
and adopt the same $r^{-1}$ slope, which is motivated by internal shock heating and is equivalent to the $\gamma = 4/3$ index of the torus model. As a fiducial normalization, we adopt $f_{\rm vir}=0.25$ for $q=1/3$ \citep{2020ApJ...895...29M}.

\subsubsection{Dust Properties and Extinction}

We adopt a highly simplified dust model, with the goal of estimating the order of magnitude significance of dust extinction in circumbinary material. We assume that dust forms when the post-shock temperatures are below a condensation temperature, $T_{\rm sh} \lesssim \Tc=10^3$~K. Below $\Tc$, we assume a constant mass fraction of material forms grains, $\Xd=5\times10^{-3}$,  typical of grain condensation in solar-composition material. We adopt a fiducial V-band opacity for these grains per gram of dust of $\kd = 10^3$~cm$^2$~g$^{-1}$.  Dust opacities depend relatively sensitively on the grain size limits, distribution, and composition. Our choice is motivated by the comparatively small grains expected to initially coalesce in the ejecta \citep{2020MNRAS.497.3166I}.  

In reality, the creation and destruction of grains takes time, and a time-dependent dust condensation, growth, and sputtering model could be used to develop much more sophisticated models of the ejecta--dust interaction \citep[e.g.][]{2016A&A...594A.108H,2022A&A...657A.109H}. For now, we note that the dust-to-gas fraction and the grain size distribution will be affected by the density and temperature of the outflow \citep[e.g.][]{2013ApJ...768..193L,2018MNRAS.478L..12G}, and that revisiting these schematic properties in the future will offer significant insight into the dust forming properties of these coalescing binary systems. Such an exercise has been recently undertaken in the limit of asymptotic expansion of simulated ejecta by \citet{2020MNRAS.497.3166I}. 

In either of our models, we integrate the total optical depth due to dust extinction along a line of sight (in the equatorial plane for the torus model),
\beq
\tau_{\rm d} = \int_{\Rs}^{\infty}\rho \Xd \kd  dr, 
\eeq
where $\rho$ is a function of $r$ in the wind model, and $r$ and $z$ in the torus model and $\Xd$ is a function of the local temperature. We then specify $A_V = 1.086 \tau_{\rm d}$.

\section{Pre-Outburst Obscuration}\label{sec:results}

In this section, we explore the dependence of dust formation and obscuration on binary system parameters. 

\subsection{Time evolution}

\begin{figure}[tbp]
\begin{center}
\includegraphics[width=0.49\textwidth]{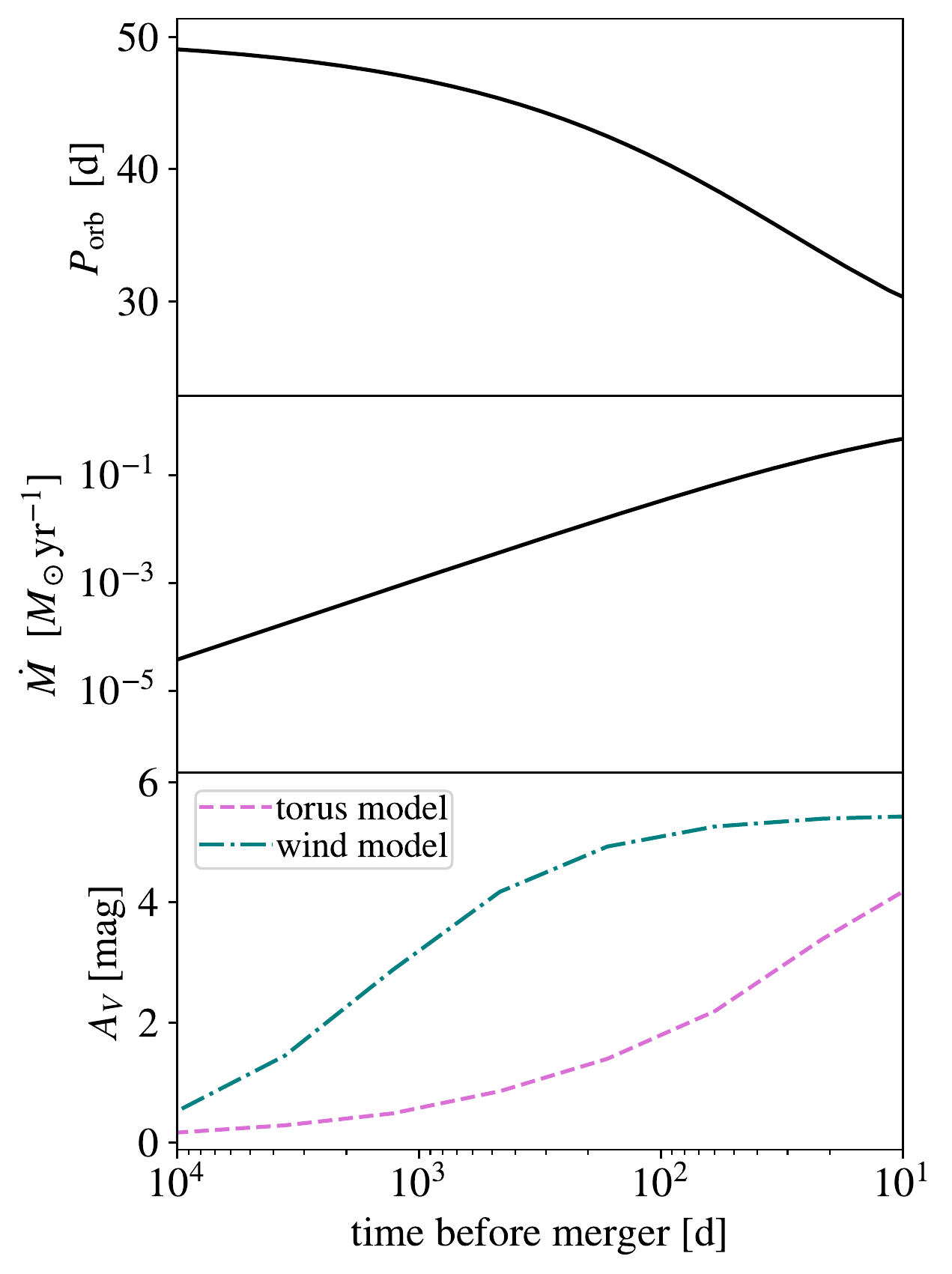}
\caption{Time evolution of orbital period, circumbinary mass loss rate, and optical obscuration, in an example merging binary system. This system consists of an $\Ms=M_\odot$, $\Rs=30R_\odot$ primary star unstably transferring mass to a $M_\odot/3$ companion. We integrate this example until the separation equals $\Rs$ and the pair merges. In both the wind and torus models of pre-outburst obscuration, the degree of obscuration increases as the mass loss intensifies and the system's merger approaches.  }
\label{fig:ts}
\end{center}
\end{figure}

We begin, in Figure \ref{fig:ts}, by showing the temporal evolution of an example system from unstable Roche lobe overflow until the orbit shrinks to the primary star's radius, $a=\Rs$. We imagine this as a point of transition from the ``lead-in" to the engulfment of the secondary object that initiates the common envelope phase itself. The system in Figure \ref{fig:ts} consists of an $\Ms=M_\odot$, $\Rs =30R_\odot$ primary star transferring mass toward a $\Ma = M_\odot/3$ companion ($q=1/3$). The system is otherwise modeled by our fiducial parameters described in Section \ref{sec:model}. 

As the binary orbital period decreases, the degree of Roche lobe overflow of the primary star  increases, driving the rate of mass loss from the system to increase also, as described by equation \eqref{mdot} and shown in the middle panel of Figure \ref{fig:ts}. This increasing mass loss rate accelerates the rate of orbital decay, as described by equation \eqref{adot}. The coupled mass loss and orbital decay lead to the eventual deposition of $\Delta m \sim 0.25 \Ma$ into the circumbinary environment  \citep{2020ApJ...895...29M}.

The lower panel of Figure \ref{fig:ts} shows the time-dependent obscuration of the merging binary due to this circumbinary material. We show results from both the torus and wind models for the circumbinary distribution, which are described in Sections \ref{sec:torus} and \ref{sec:wind}, respectively. In the torus model, we adopt a viewing angle through the torus midplane (which is equivalent to the equatorial plane of the binary system). In both models, the dust obscuration, $A_V$, increases from very low values long before merger to higher values at later times, reflecting the increasingly mass-rich circumbinary environment of the binary. The wind model exhibits higher obscuration than the torus model, especially at early times. This difference represents the greater radial extent of the unbound wind relative to the bound torus. Within the final $10^2$~d before the binary merges, the wind model levels out as newly-lost material does not have time to propagate outward to the dust condensation radius. Within the hydrostatic torus model, the circumbinary material is assumed to instantaneously assume its hydrostatic configuration.  The timescale for this adjustment is not known in detail, but it is likely at least the stellar dynamical time. Therefore, the predicted obscuration in the last orbit or perhaps several orbits might not be realistically modeled in the torus case because this delay for hydrostatic adjustment is not explicitly accounted for.  In Figure \ref{fig:ts}, this uncertainty would correspond to the last $\sim 30$~d, for example. The immediate pre-merger behavior thus represents a factor $\sim 2$ uncertainty in the torus model. 

Despite these differences, it is interesting to note that both models reach a similar estimate of several magnitudes of dust extinction at the time of merger. Viewed over time under this increasing obscuration, we might see an optical precursor of the outburst to fade while its infrared counterpart brightens. This evolution points to the leverage that detailed pre-outburst observations of stellar-coalescence transients can provide on these sorts of models of circumbinary mass loss  \citep{2011A&A...528A.114T,2017ApJ...834..107B,2020MNRAS.496.5503B,2021A&A...653A.134B,2021A&A...646A.119P}.

\subsection{Dependence on binary properties}

\begin{figure}[tbp]
\begin{center}
\includegraphics[width=0.49\textwidth]{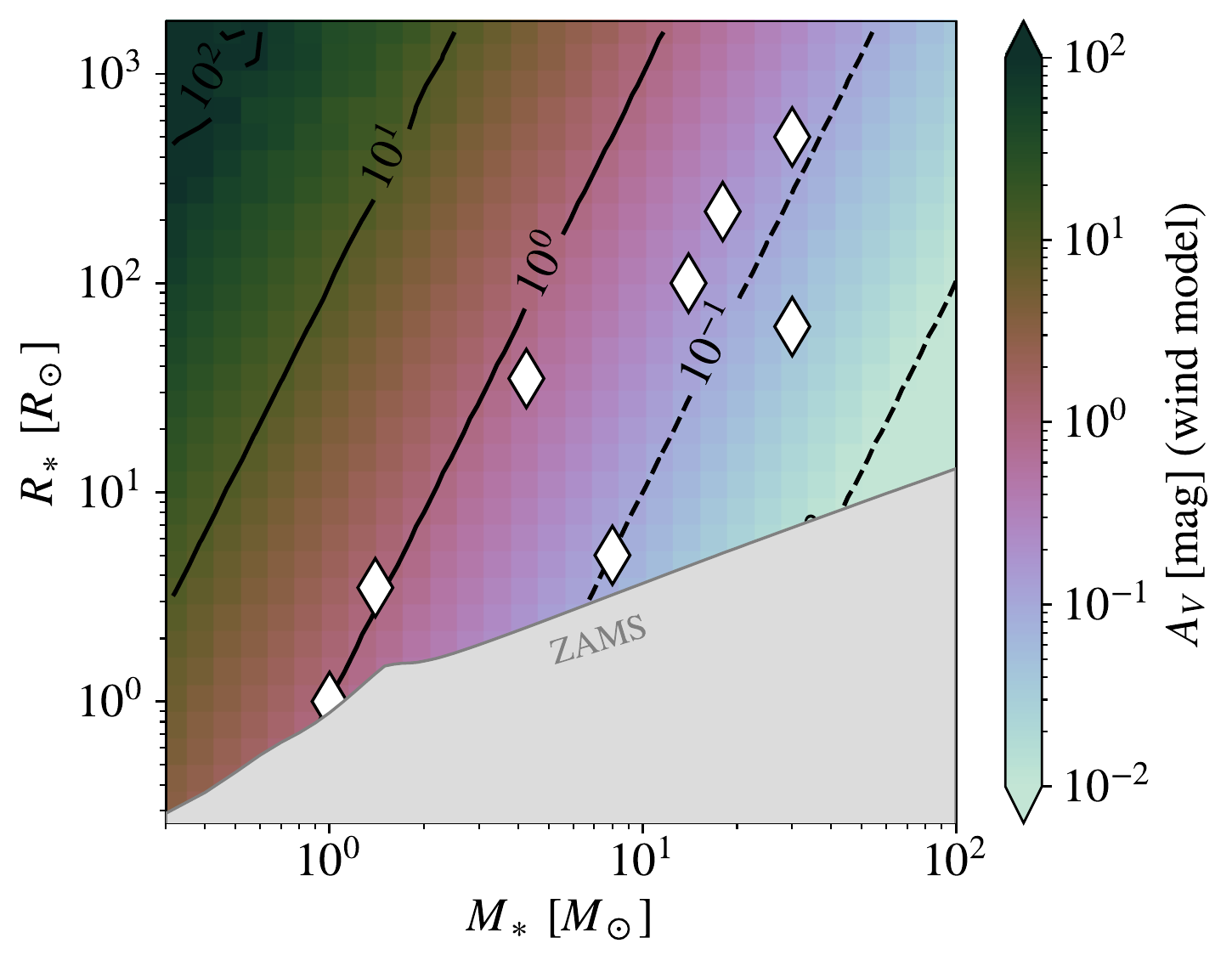}
\includegraphics[width=0.49\textwidth]{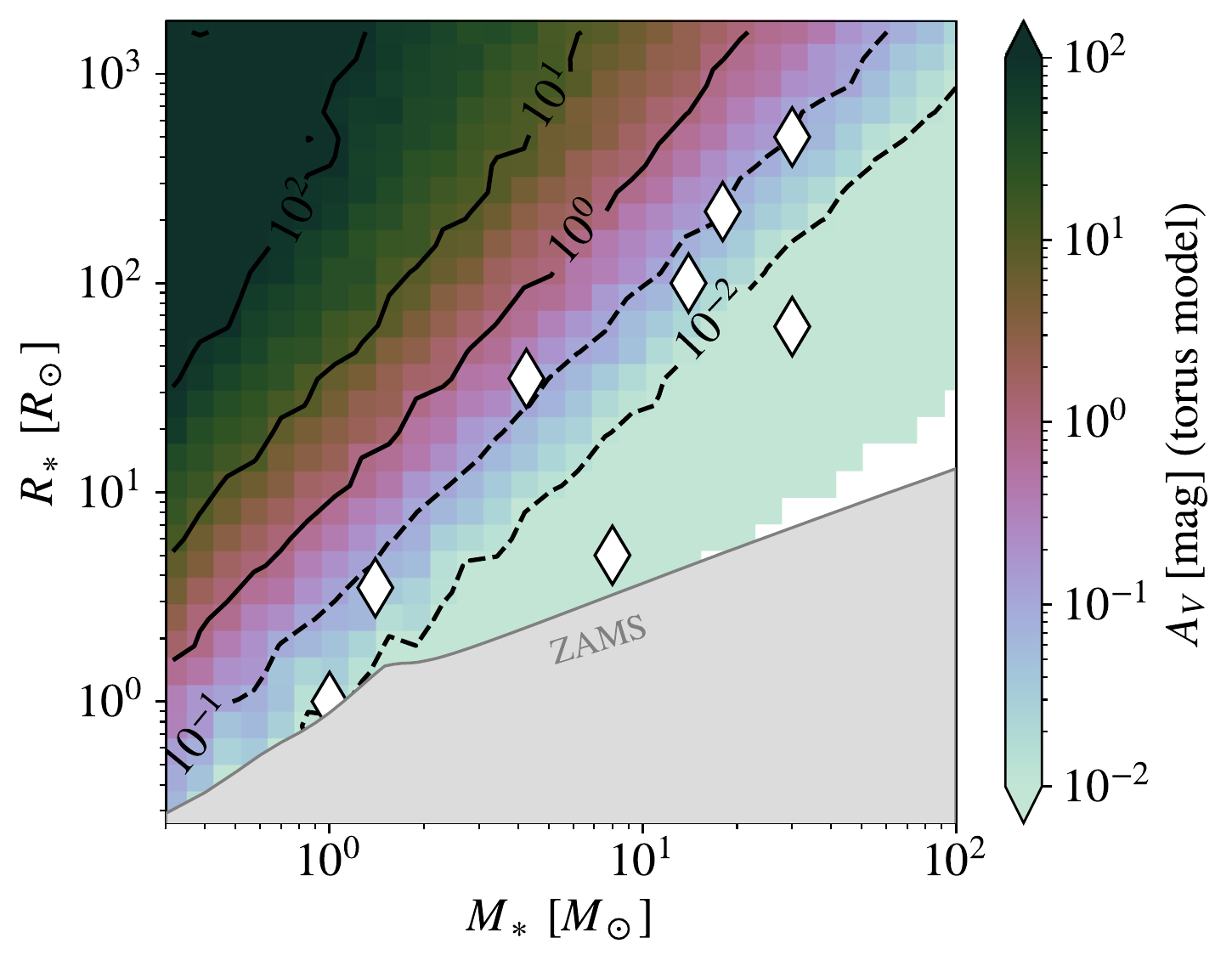}
\caption{Predicted obscuration just prior to merger in the wind and torus  (equatorial line of sight) models given primary star mass and radius. Each model assumes $q=1/3$. The progenitor properties of known LRNe  are marked with diamonds. In both models, low mass, extended radius primary stars should lead to highly dust-obscured transients, in some cases reaching $A_V \gg 10$. By contrast, more compact or massive primary stars should lead to low levels of pre-outburst obscuration. The observed population of optical LRNe  traces predicted $A_V \lesssim 1$ in either model.   }
\label{fig:mr}
\end{center}
\end{figure}

In Figure \ref{fig:mr}, we map the maximal $A_V$ measured at the time of merger ($a=\Rs$) in our wind and torus models. Our results show that this maximal absorption depends on both stellar mass and radius. At the simplest level this result can be traced to the characteristic temperatures of circumbinary outflows. As material is shock-heated and expelled, the post-shock temperatures depend on the compactness of the primary star. More massive, smaller radius primary stars represent deeper gravitational wells and will have hotter, faster moving circumbinary material, that is, by consequence, less dust-forming. The cool circumbinary surroundings of low mass, extended radius stars become dramatically obscuring, perhaps providing tens of magnitudes of extinction in the most extreme cases. 

The qualitative trends and normalization of the wind and torus model predictions are similar. However, Figure \ref{fig:mr} does highlight that these models exhibit somewhat different parameter dependence. We empirically derive the following relations based on models in which we randomly varied model parameters. For the wind model, 
\begin{align}\label{approxwind}
A_{V, {\rm wind}} \approx 0.97 &\left(q \over 0.1 \right)^{1.04}  \left( \Ms \over M_\odot \right)^{-1.5} \left(\Rs \over 10\right)^{0.5} \nonumber \\ 
\times &\left( \Xd \kd \over 5~{\rm cm}^2~{\rm g}^{-1}\right) \left(f_{\rm vir} \over 0.25 \right)^{-2.5} \left(f_v \over 0.25 \right)^{0.5},
\end{align}
while for the torus model with an equatorial line of sight, 
\begin{align}\label{approxtorus}
A_{V,{\rm torus}} \approx 0.18 &\left(q \over 0.1 \right)^{1.6}  \left( \Ms \over M_\odot \right)^{-2.8} \left(\Rs \over 10\right)^{1.8} \nonumber \\ 
\times &\left( \Xd \kd \over 5~{\rm cm}^2~{\rm g}^{-1}\right). 
\end{align}

In this context, it is worth highlighting that our simple assumption of unchanging dust physics across the entire parameter space neglects the likely dependence of dust formation on the density and velocity of ejecta. In cases of very high predicted extinction, e.g. $A_V\sim 10^2$, larger grains might instead form, impacting the opacity as a function of wavelength and the infrared appearance of obscured transients. Equation \eqref{approxtorus} only parameterizes the obscuration along the equatorial line of sight. We find that within the torus model there is viewing angle dependence: at polar angles, $\cos(i)\lesssim0.4$, the sight line to the central binary is unobscured while at more equatorial angles, $\cos(i)\gtrsim 0.4$, $A_V$ is approximately constant.  

\begin{figure}[tbp]
\begin{center}
\includegraphics[width=0.49\textwidth]{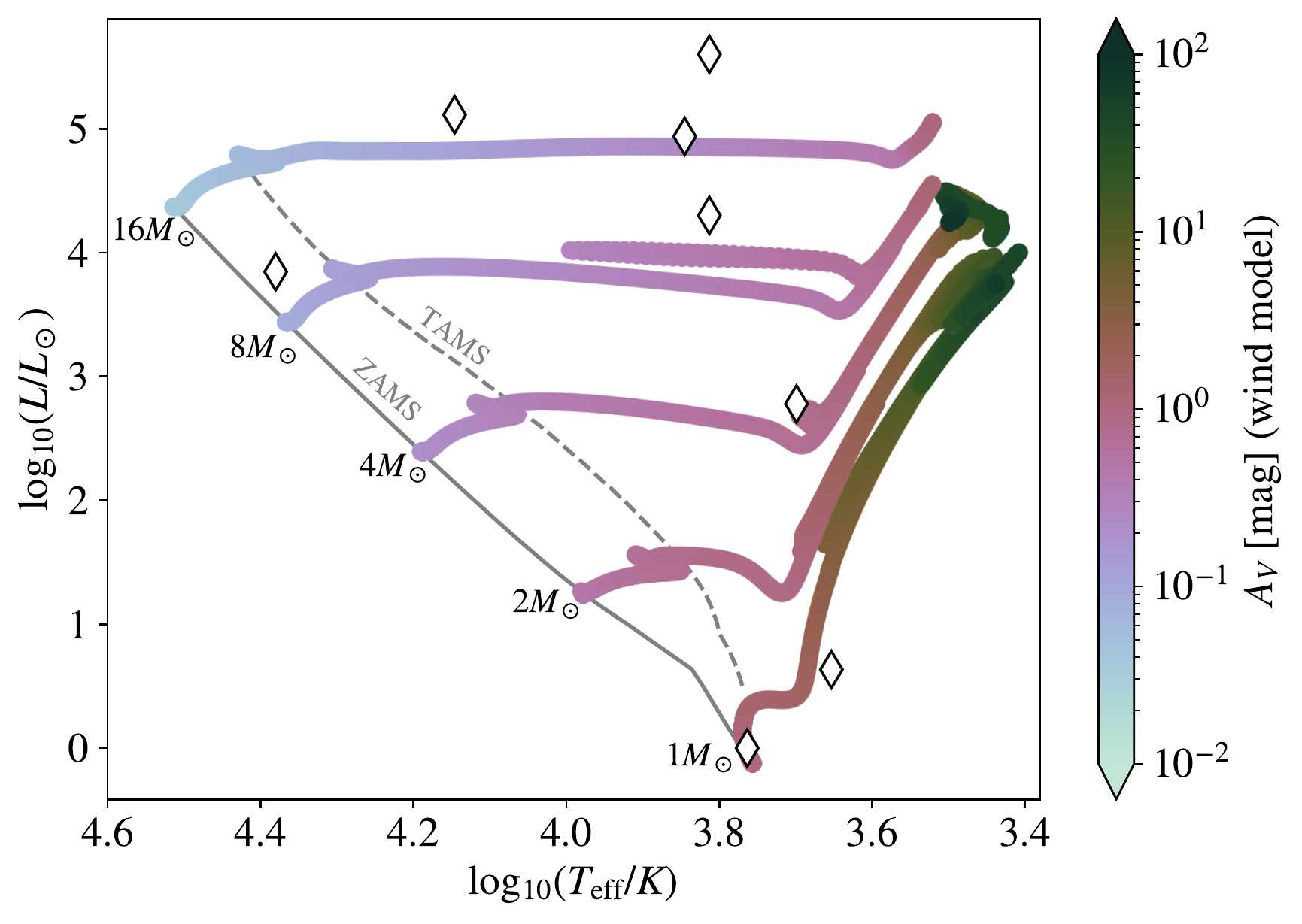}
\includegraphics[width=0.49\textwidth]{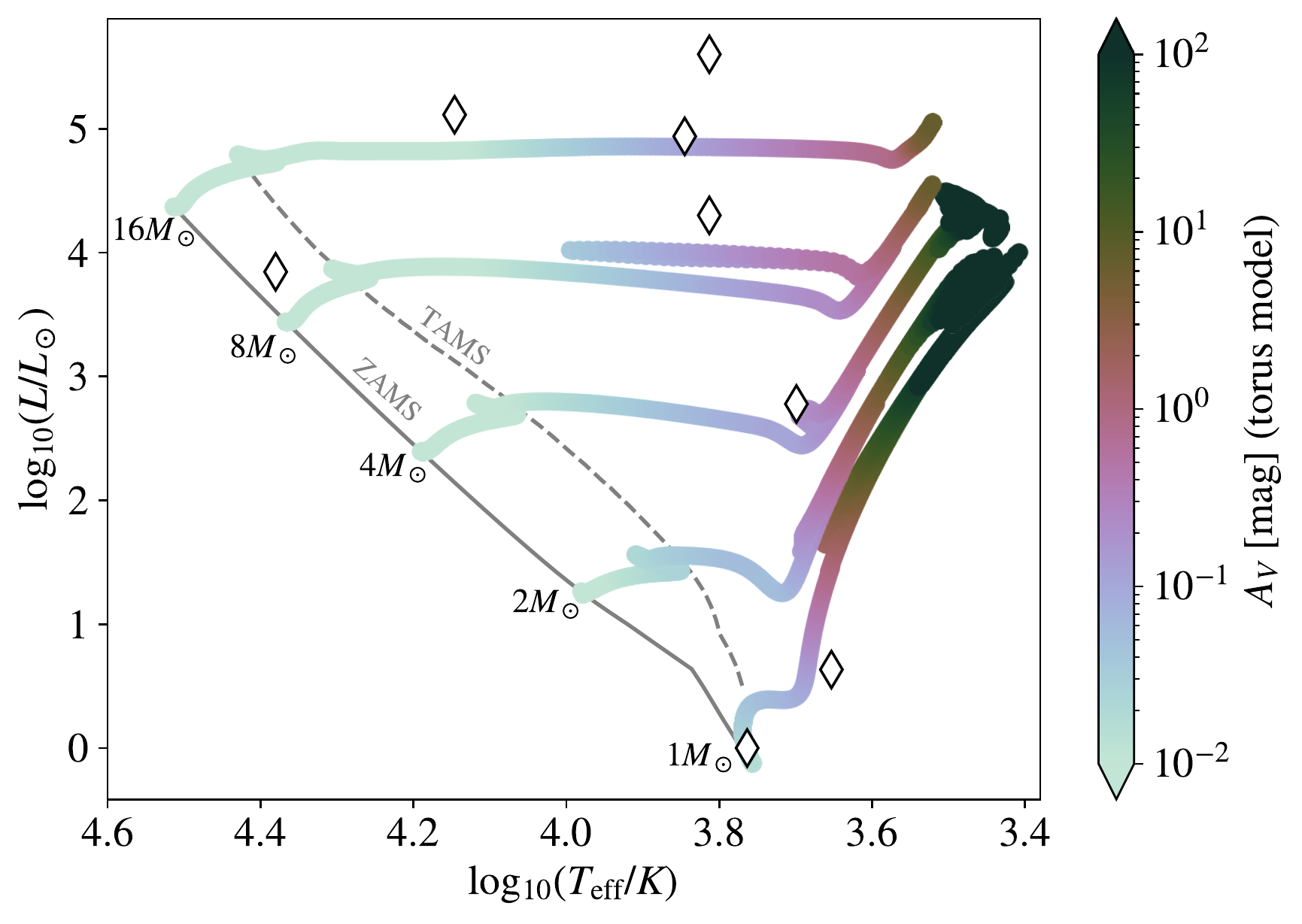}
\caption{Predicted obscuration in the context of the HRD. This figure adopts wind and torus models with $q=1/3$. The primary star properties are realized from MIST stellar tracks of solar metalicity. The lower-mass extended radii primary stars that lead to the highest predicted $A_V$ occur on the giant branch of primary stars with $\Ms \lesssim 4M_\odot$. These systems have low $T_{\rm eff}\lesssim 4000$~K, and relatively high luminosities $L\gtrsim 10^2 L_\odot$.  By contrast, the progenitors of optical LRNe have a wide range in mass, but a narrow range in effective temperature, primarily being mildly evolved yellow giants crossing the Hertzsprung gap, properties that correspond to low levels of pre-outburst obscuration. }
\label{fig:hr}
\end{center}
\end{figure}

Primary star mass and radius thus clearly affect the degree of obscuration due to circumbinary mass loss. These primary star properties also correlate with different stellar evolutionary states and appearances. Figure \ref{fig:hr} traces circumbinary obscuration in the wind and torus models in the Hertzsprung-Russell Diagram (HRD). We show evolutionary tracks of 1, 2, 4, 8, and 16$M_\odot$ stars at solar metallicity from MIST \citep{2016ApJ...823..102C}. Here, we can immediately observe that merger events on the main sequence will be largely unobscured, with $A_V\ll1$~mag. Similarly, mergers of Hertzsprung gap objects, with $T_{\rm eff} \gtrsim 5500$~K, also experience low degrees of obscuration. 

This can be contrasted to the late evolutionary phases of low-mass stars. In the tracks of the 1, 2, and $4M_\odot$ stars, very high degrees of dust formation and obscuration $A_V \gg 5$~mag are observed near the tips of their respective giant branches. This location in the HRD corresponds with low effective temperatures, $T_{\rm eff}\lesssim 4000$~K, and high luminosities $L\gtrsim10^2L_\odot$. Thus, before dust obscures these merging systems, they are preferentially old, luminous, and cool giants.

\begin{figure}[tbp]
\begin{center}
\includegraphics[width=0.49\textwidth]{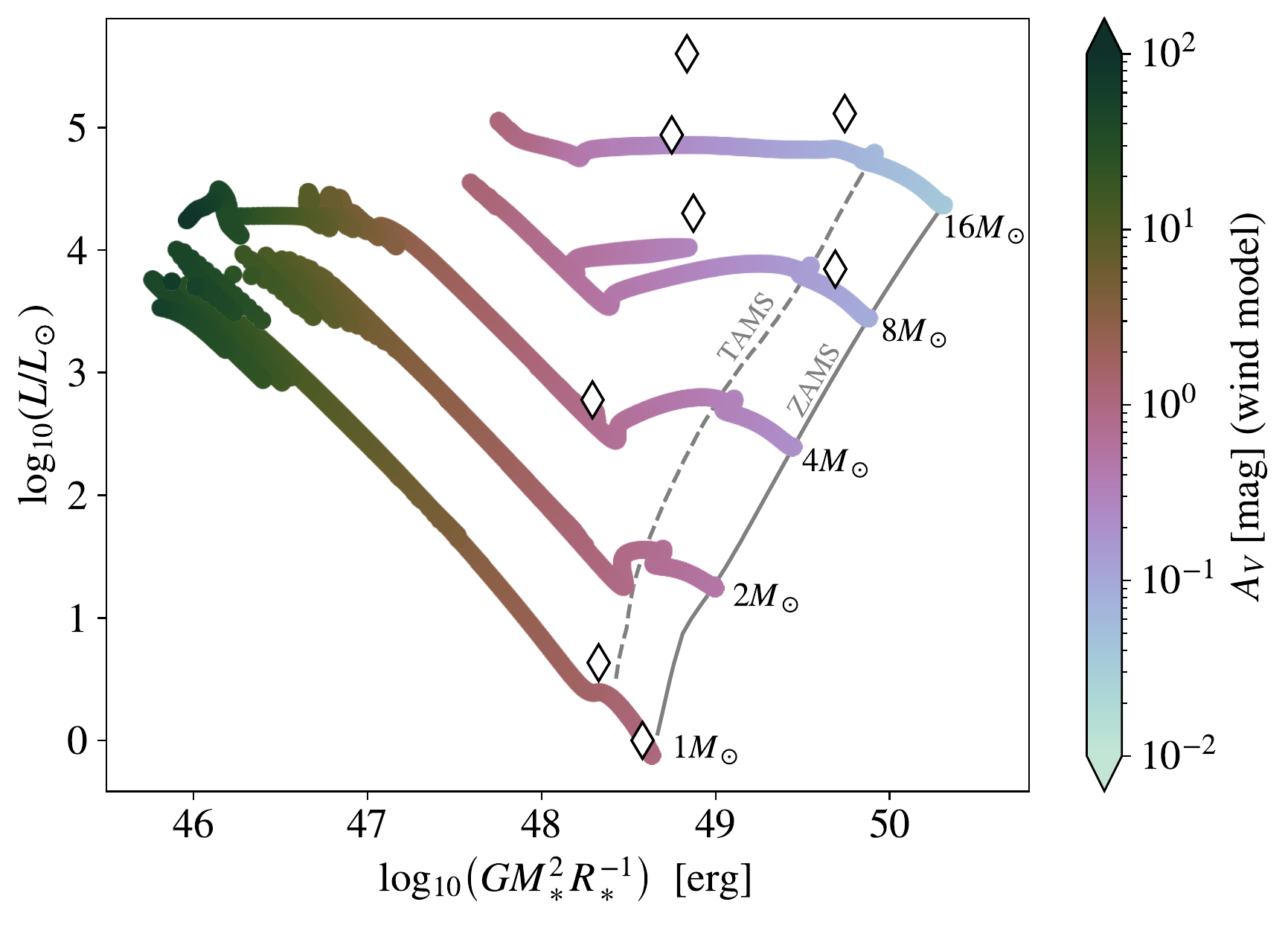}
\includegraphics[width=0.49\textwidth]{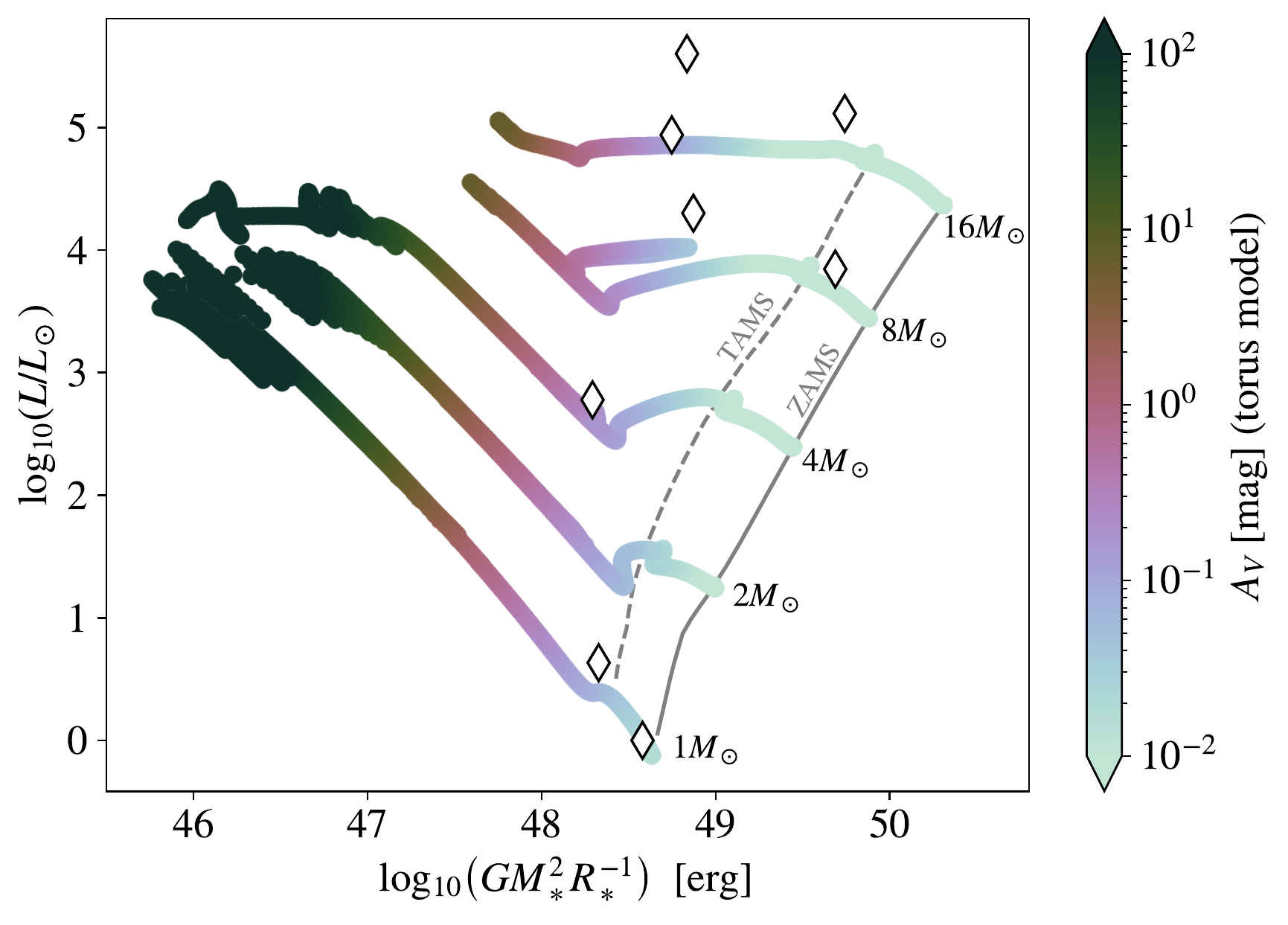}
\caption{Predicted obscuration in the context of primary star binding energy and luminosity; otherwise identical to Figure \ref{fig:hr}. Low-binding-energy primary stars are those that are more likely to lead to successful common envelope ejection in interactions with a companion object. We find that low binding energy correlates with predictions of highly obscured mergers, while higher binding energies have low predicted $A_V$. The population of optical LRNe traces the low-obscuration, high binding energy phase space of merger outcomes.   }
\label{fig:hrE}
\end{center}
\end{figure}

Figure \ref{fig:hrE} shows these same stellar tracks in the context of a common-envelope HRD, where we show the binding energy of primary stars along with their luminosity. Common envelope ejection is believed to be related to the overall system energy balance, in which to unbind a common envelope $\Delta E \sim E_{\rm bind}$, where $E_{\rm bind}\sim G\Ms^2 / \Rs$ is the binding energy of the primary star's envelope before the coalescence \citep[e.g.][]{1984ApJ...277..355W}. The energy source, $\Delta E$, is largely the gravitational energy of the decaying companion star orbit, $\Delta E \sim E_{\rm final}\sim G \Ms^2 q / a_{\rm final}$. At the crudest level, therefore, envelope ejection requires $a_{\rm final} / \Rs \lesssim  q$. In more detailed analysis, coefficients can change these scalings by an order of magnitude \citep[e.g.][]{1984ApJ...277..355W,1993PASP..105.1373I,2011MNRAS.411.2277D,2013A&ARv..21...59I}. 

In Figure \ref{fig:hrE}, we see that systems with lower binding energies that will tend to lead to common envelope ejection outcomes are predicted to be highly obscured. Higher binding energy systems -- those closer to their main sequence radii -- tend to lead to low predicted obscuration and are more likely to correspond to merger outcomes.

\subsection{The progenitors of Luminous Red Novae}

We compare the population of optical transient progenitors (Section \ref{sec:lrn}) to our model predictions for degree of obscuration. In Figure \ref{fig:mr}, we plot the LRNe progenitors on the primary-star mass--radius plane, and in Figure \ref{fig:hr}, we plot these progenitors on the HRD.  Both the mass--radius and HRD views show that LRNe progenitors lie in phase space where we expect relatively low degrees of pre-outburst dust extinction. In the torus model, systems typically exhibit $A_V \lesssim 0.1$~mag, while in the wind model $A_V \lesssim 1$~mag. In either case, these predictions are consistent with a largely unobscured transient. We note that an unobscured transient is always possible for a pole-on line of sight in the torus model, but this being a viewing-angle effect is unlikely given that V1309 Sco is known to be viewed in the equatorial plane. 

In Figure \ref{fig:hrE}, we see that the lower binding energy primary stars are systematically not observed in the present LRNe sample. This is highly suggestive that the unobscured, optical LRNe sample traces merger outcomes, while dusty, obscured transients will characterize common envelope ejections. 

There is some evidence in precursor lightcurves of LRNe for some degree of time-variable extinction. V1309 Sco and M31 LRN 2015 each exhibit a slow rise of precursor brightening followed by a optical fade ($\sim 1$~mag) just before the optical outburst \citep{2011A&A...528A.114T,2020MNRAS.496.5503B}. If these dips were due to time-dependent dust reddening rather than a bolometric fade that would be consistent with the pre-outburst mass loss model. Indeed, the presence of infrared excess in the pre-outburst phase of V1309 Sco is suggestive that this is the case \citep{2016A&A...592A.134T}. For V1309 Sco ($\Ms\sim 1.5 M_\odot, \ \Rs\sim3.5R_\odot$), the predicted $A_V$ is 1~mag in the wind model and 0.1~mag in the torus model. For M31 LRN 2015 ($\Ms\sim 4 M_\odot, \ \Rs\sim30R_\odot$), the model predictions are very similar, suggesting that the mild fade observed pre-outburst is entirely consistent with dust reddening.

\subsection{Dusty transients in the binary population}

\begin{figure}[tbp]
\begin{center}
\includegraphics[width=0.49\textwidth]{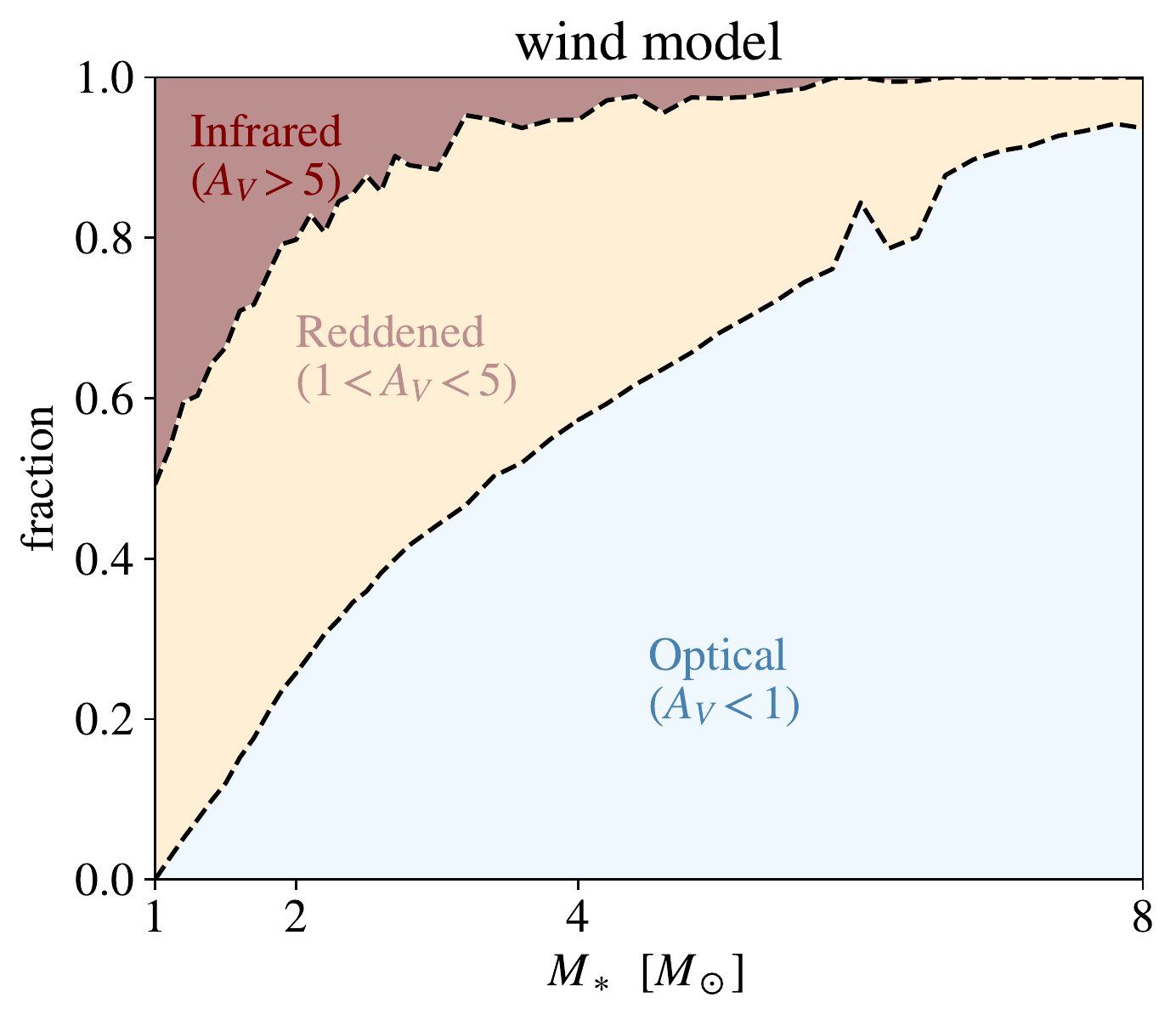}
\includegraphics[width=0.49\textwidth]{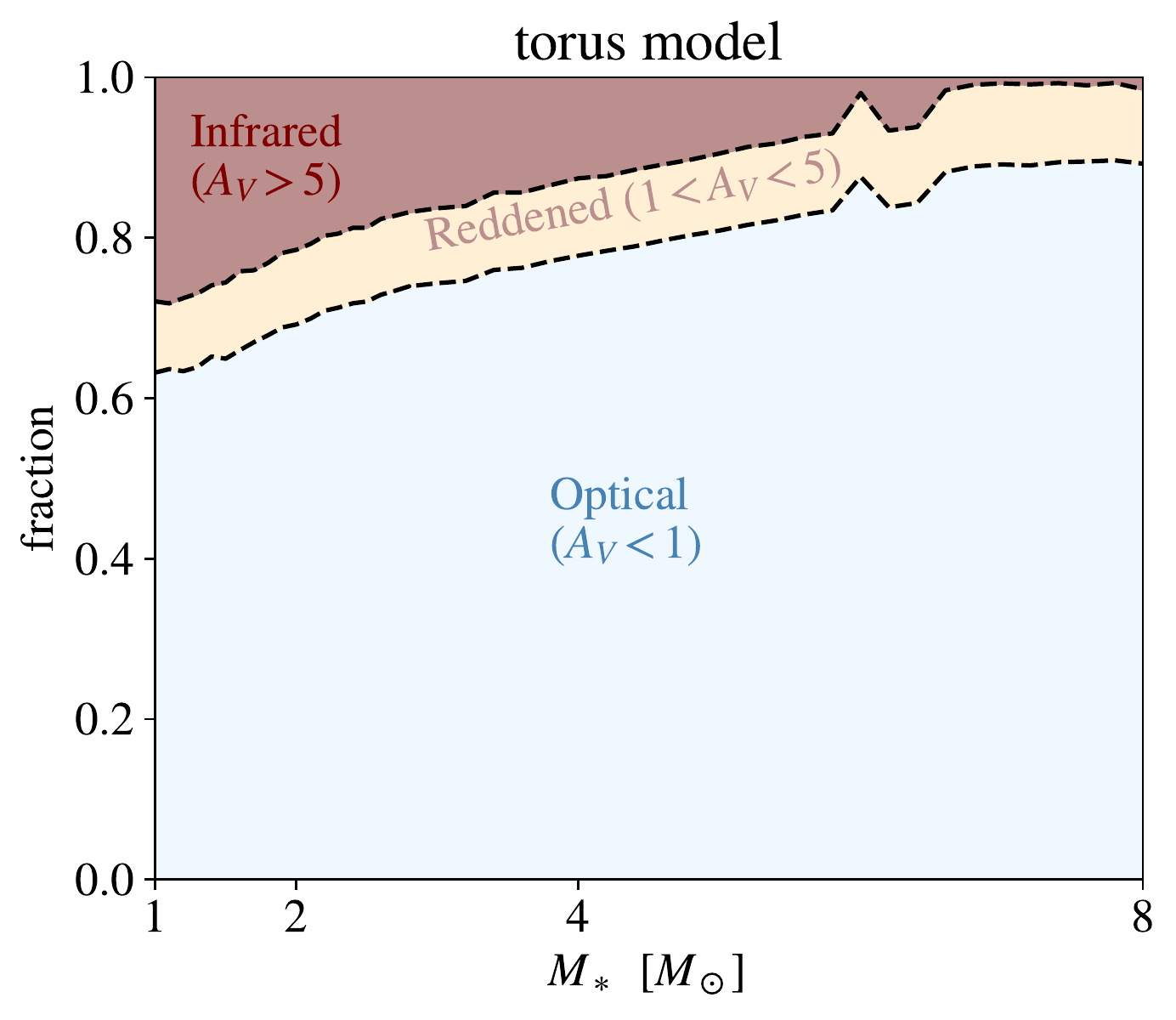}
\caption{Relative population of merger transients as a function of primary star mass. In this model, we assume all mergers have $q=1/3$. We compute the predicted $A_V$ along MIST solar metalicity tracks, and show the fractions of merging systems with low, intermediate, and high $A_V$, assuming a logarithmic merging separation distribution. Among approximately solar mass stars, on the order of half of transients will be highly dust-obscured, primarily-infrared transients. As we consider higher mass systems, a greater fraction will be primarily-optical transients.   }
\label{fig:frac}
\end{center}
\end{figure}

We have argued that LRNe originate from the relatively unobscured population of binary merger events. Their more obscured counterparts will be dusty and optically faint, while perhaps being infrared luminous. We imagine that these sources could have extremely red colors of the SPRITEs class of exclusively-infrared transients reported by \citet{2019ApJ...886...40J} from the Spitzer Infrared Intensive Transients Survey \citep[SPIRITS,][]{2017ApJ...839...88K}.

To understand the imprint of dusty transients on the population of binary merger and common-envelope transients, we make a simple estimate of their fractional contribution to the overall merging population. 
To do so, we assume that binaries are distributed with a broad separation distribution with equal number of binary systems per logarithmic bin in separation \citep{1924PTarO..25f...1O}. That implies that as stellar radii grow, the probability of merging with a companion per log radius is constant. In Figure \ref{fig:frac}, we trace MIST evolutionary tracks from the main sequence out the giant branch, assuming an equal number of mergers per log radius of the star. We apply equations \eqref{approxwind} and \eqref{approxtorus} to estimate $A_V$ in these events.  We assume isotropic viewing angles. In the spherical wind case of equation \eqref{approxwind} this has no effect. However, for the torus, we assume 40\% of viewing angles are sufficiently pole-on to have $A_V=0$, otherwise we apply equation \eqref{approxtorus}. 
Finally, we divide these events into groups of low, intermediate, and high levels of obscuration. The most obscured systems will be those where the optical signature is nearly completely extinguished, and will manifest as infrared transients. The intermediate levels of reddening will lead to highly reddened optical and infrared transients, while the low extinction case will lead to primarily-optical transients.

Figure \ref{fig:frac} shows that at $1M_\odot$, about 30--50\% of mergers lead to infrared transients. This fraction decreases to nearly zero above $8M_\odot$.  We can understand this high fraction of infrared transients from low-mass systems in context of the observed events in two ways. First, we consider a volume-limited sample of stellar coalescence events, such as all the events occurring within the galaxy. These galactic events will be greatly dominated (by number) by low-mass events, with masses similar to $1M_\odot$, because of the steeply-decreasing stellar initial mass function (IMF) $dN/d\Ms \propto \Ms^{-2.3}$. We note that this is true despite the increasing fraction of massive stars in interacting binaries, which partially counteracts the IMF. The fraction of interacting systems is about 15\% for solar-mass primary stars, but increases to nearly 100\% for O-type stars, with the interacting binary fraction scaling roughly as $f_{\rm interact}\propto\Ms^{0.7}$ \citep[See Figure 42 of][]{2017ApJS..230...15M}, thus $f_{\rm interact} dN/d\Ms \propto \Ms^{-1.6}$.  Within such a volume limited population (weighted by $\Ms^{-1.6}$), in the wind model, 25\% of events have $A_V<1$, 48\% have $1<A_V<5$, and 26\% have $A_V>5$. For the same weighting  in the torus model, 69\% of events have $A_V<1$, 9\% have $1<A_V<5$, and 22\% have $A_V>5$. 

Given the nature of these sources as astronomical transients, we might also consider a luminosity limited sample. \citet{2014MNRAS.443.1319K} has shown that empirically, $L_{\rm LRN}\propto \Ms^3$. This scaling more than cancels the slope of the IMF and interacting binary fraction, implying a similar number of detected LRNe across all masses, with a possible weighting toward more massive stars. While the number of sources remains small, this is at least partially born out by the broad mass distribution of the observed sample. In this luminosity-weighted sample, optical events will dominate because of their higher luminosities and the greater contribution from massive stars. 

Finally, it is important to consider that dust-enshrouded, infrared transients are associated with different primary stars than optical counterparts. These progenitor stars tend to be cool, luminous, extended companions with lower binding energies (e.g. equations \eqref{approxwind} and \eqref{approxtorus} and Figures \ref{fig:hr}, and \ref{fig:hrE}). The correlation of these properties with possible common envelope ejection outcomes suggests that common envelope ejection transients will systematically be among the obscured, infrared population. The higher binding energy envelopes associated with merger outcomes will lead to less obscuration and  optical transients.

\subsection{OGLE-2002-BLG360: A Prototype Dusty Merger Transient?}\label{sec:blg360}

The transient OGLE-2002-BLG360, originally thought to be a lensing event, has been interpreted as a possible LRN transient \citep{2013A&A...555A..16T}. The system gradually brightened into a nearly thousand-day outburst in 2002. The progenitor was a $T_{\rm eff}\sim 4300$~K giant. The distance was uncertain, but adopting a distance of $8.2$~kpc, \citet{2013A&A...555A..16T} derive $\Rs \sim 30 R_\odot$ and $L \sim 300L_\odot$. The old galactic bulge stellar population implies a low progenitor mass, perhaps $\Ms \sim M_\odot$.  Intriguingly, \citet{2013A&A...555A..16T} find that the pre-outburst spectral energy distribution requires an extinction of $A_V \sim 3$ by warm $\sim 800$~K dust, which they note is only consistent with ongoing dust formation in a continuous outflow. As the source evolved, it eventually faded in the optical as it became so dust-obscured that $A_V \gtrsim 20$ is required to accommodate the constraints. 

Our models predict $A_V \sim $3--10 for a 1--2$M_\odot$, approximately 30$R_\odot$ primary star. The long duration of the transient is likely related to the long orbital period about the extended primary, 
\beq
P_{\rm orb}(a) \sim 19~{\rm d} \left( \frac{\Ms}{M_\odot} \right)^{-1/2} \left( \frac{a}{30 R_\odot} \right)^{3/2}. 
\eeq
For comparison, the expansion time of material to the dust condensation radius is,
\beq
t_{\rm dust} \sim 33~{\rm d}  \left( \frac{f_{\rm vir}}{0.02} \right)  \left( \frac{\Ms}{M_\odot} \right)^{1/2}  \left( \frac{\Rs}{30 R_\odot} \right)^{1/2} 
\eeq
if we assume that $r_{\rm dust}$ is given by equation \eqref{rdust} and $v_{\rm ej} \sim (2G\Ms/\Rs)^{1/2}$. The similarity of these timescales is indicative that the ejecta expansion and cooling is ongoing even during the outburst phase of the event. Thus, the entire transient episode is similar to the wind-like pre-outburst behavior where the temporal evolution is dictated by the changing mass outflow. This is different from the other LRNe, in which the duration of the event is dictated by the expansion time of the ejecta, because $t_{\rm dust} \gg P_{\rm orb}$.
This is indicative that dusty transients tend to be more wind-like as opposed to impulsive, and that the duration of the events will be set by the duration of mass ejection rather than material expansion. 

We conclude that OGLE-2002-BLG360 is very likely a prototype event for an intermediate level of dust obscuration, positioned in the phase space of reddened transients in Figure \ref{fig:frac}. We anticipate that more detailed modeling of the abundant data for this source will be very informative.

\section{Summary and Conclusions}\label{sec:conclusions}

In this paper, we have analyzed the thermodynamic evolution of the ejecta of stellar coalescence transients, and of the mass lost leading up to these episodes. When dust condenses in this circumstellar material, it dramatically increases the opacity obscures optical light from the merging system while redistributing emission toward the infrared. We find that the degree of this obscuration depends on the properties of the merging system. Some key results of our study are:

\begin{enumerate}
\item After being shock-heated to a few percent the Virial temperature, the ejecta in LRNe appear to decline linearly with radius, proportional to the gravitational potential. This implies continuous heat injection by internal shocks are crucial in regulating the ejecta thermodynamics  (Figure \ref{fig:r90}). 
\item Pre-outburst mass loss accompanies orbital decay in merging binary systems. The resulting circumstellar material can lead to time-dependent obscuration or even the optical vanishing of coalescing systems due to dust formation (Figure \ref{fig:ts}). 
\item The most obscured systems are those with lower primary-star masses and extended radii high on the giant branch, while more compact or massive systems tend to be comparatively unobscured (Figures \ref{fig:mr} and \ref{fig:hr}). 
\item Because lower envelope binding energy is correlated with higher $A_V$, the systems most likely to lead to common envelope ejection will tend to systematically produce dusty, infrared transients (Figure \ref{fig:hrE}). 
\item The population of optical LRNe are all associated with model predictions of low obscuration (Figures \ref{fig:mr}, \ref{fig:hr}, and \ref{fig:hrE}). A corresponding population of dusty, optically-obscured, infrared-luminous transients arise from the $\sim$25\% of stellar coalescence events involving lower-mass extended donor stars (Figure \ref{fig:frac}). 
\item OGLE-2002-BLG360, with properties slightly distinct from many of the optical LRNe, may be a prototype of a partially dust obscured transient intermediate between the optical and fully infrared events. 
\end{enumerate}

The fact that dusty, infrared transients should exist, and that their properties are correlated with the properties of the donor star in the binary system, indicates the breadth of the population of transients related to stellar coalescence. The optically-luminous LRNe appear to be just one subset of this full class of events, with properties that suggest they tend to be associated with stellar merger (rather than common envelope ejection) outcomes. More broadly exploring this parameter space, both in modeling and observationally, is a promising path toward a deeper understanding of these transformative events in binary stellar evolution. With time-domain capabilities expanding in both the optical and, more recently, infrared \citep[e.g.][]{2020PASP..132b5001D}, we are well positioned to begin mapping the range of possible events. To this end, the anticipated near-infrared follow-up capabilities of the {\it James Webb Space Telescope} will likely be pivotal in revealing the properties of these dusty systems.

\section*{Reproduction Software and Data}
Software and data to reproduce the results of this work are available via github at: \url{https://github.com/morganemacleod/dustytransients_materials} \citep{morgan_macleod_2022_6554801}. 

\acknowledgements
We thank Nadia Blagorodnova, Tomasz Kami\'nski, and Jonathan Grindlay
for helpful conversations that led to and informed this work.
This work was supported by the National Science Foundation under Grant No. 1909203.  Support for this work was provided by NASA through the NASA Hubble Fellowship grant \#HST-HF2-51477.001 awarded by the Space Telescope Science Institute, which is operated by the Association of Universities for Research in Astronomy, Inc., for NASA, under contract NAS5-26555.

\software{IPython \citep{PER-GRA:2007}; SciPy \citep{2020SciPy-NMeth};  NumPy \citep{van2011numpy};  matplotlib \citep{Hunter:2007}; Astropy \citep{2013A&A...558A..33A}; RLOF \citep{RLOF1.1}; reproduction materials for this work \citep{morgan_macleod_2022_6554801}.}


\begin{thebibliography}{}
\expandafter\ifx\csname natexlab\endcsname\relax\def\natexlab#1{#1}\fi
\providecommand{\url}[1]{\href{#1}{#1}}
\providecommand{\dodoi}[1]{doi:~\href{http://doi.org/#1}{\nolinkurl{#1}}}
\providecommand{\doeprint}[1]{\href{http://ascl.net/#1}{\nolinkurl{http://ascl.net/#1}}}
\providecommand{\doarXiv}[1]{\href{https://arxiv.org/abs/#1}{\nolinkurl{https://arxiv.org/abs/#1}}}

\bibitem[{{Astropy Collaboration} {et~al.}(2013){Astropy Collaboration},
  {Robitaille}, {Tollerud}, {Greenfield}, {Droettboom}, {Bray}, {Aldcroft},
  {Davis}, {Ginsburg}, {Price-Whelan}, {Kerzendorf}, {Conley}, {Crighton},
  {Barbary}, {Muna}, {Ferguson}, {Grollier}, {Parikh}, {Nair}, {Unther},
  {Deil}, {Woillez}, {Conseil}, {Kramer}, {Turner}, {Singer}, {Fox}, {Weaver},
  {Zabalza}, {Edwards}, {Azalee Bostroem}, {Burke}, {Casey}, {Crawford},
  {Dencheva}, {Ely}, {Jenness}, {Labrie}, {Lim}, {Pierfederici}, {Pontzen},
  {Ptak}, {Refsdal}, {Servillat}, \& {Streicher}}]{2013A&A...558A..33A}
{Astropy Collaboration}, {Robitaille}, T.~P., {Tollerud}, E.~J., {et~al.} 2013,
  \aap, 558, A33, \dodoi{10.1051/0004-6361/201322068}

\bibitem[{{Banerjee} \& {Ashok}(2002)}]{2002A&A...395..161B}
{Banerjee}, D.~P.~K., \& {Ashok}, N.~M. 2002, \aap, 395, 161,
  \dodoi{10.1051/0004-6361:20021243}

\bibitem[{{Banerjee} \& {Ashok}(2004)}]{2004ApJ...604L..57B}
{Banerjee}, D. P.~K., \& {Ashok}, N.~M. 2004, \apjl, 604, L57,
  \dodoi{10.1086/383264}

\bibitem[{{Banerjee} {et~al.}(2003){Banerjee}, {Varricatt}, {Ashok}, \&
  {Launila}}]{2003ApJ...598L..31B}
{Banerjee}, D. P.~K., {Varricatt}, W.~P., {Ashok}, N.~M., \& {Launila}, O.
  2003, \apjl, 598, L31, \dodoi{10.1086/380389}

\bibitem[{{Blagorodnova} {et~al.}(2017){Blagorodnova}, {Kotak}, {Polshaw},
  {Kasliwal}, {Cao}, {Cody}, {Doran}, {Elias-Rosa}, {Fraser}, {Fremling},
  {Gonzalez-Fernandez}, {Harmanen}, {Jencson}, {Kankare}, {Kudritzki},
  {Kulkarni}, {Magnier}, {Manulis}, {Masci}, {Mattila}, {Nugent}, {Ochner},
  {Pastorello}, {Reynolds}, {Smith}, {Sollerman}, {Taddia}, {Terreran},
  {Tomasella}, {Turatto}, {Vreeswijk}, {Wozniak}, \&
  {Zaggia}}]{2017ApJ...834..107B}
{Blagorodnova}, N., {Kotak}, R., {Polshaw}, J., {et~al.} 2017, \apj, 834, 107,
  \dodoi{10.3847/1538-4357/834/2/107}

\bibitem[{{Blagorodnova} {et~al.}(2020){Blagorodnova}, {Karambelkar}, {Adams},
  {Kasliwal}, {Kochanek}, {Dong}, {Campbell}, {Hodgkin}, {Jencson},
  {Johansson}, {Koz{\l}owski}, {Laher}, {Masci}, {Nugent}, \&
  {Rebbapragada}}]{2020MNRAS.496.5503B}
{Blagorodnova}, N., {Karambelkar}, V., {Adams}, S.~M., {et~al.} 2020, \mnras,
  496, 5503, \dodoi{10.1093/mnras/staa1872}

\bibitem[{{Blagorodnova} {et~al.}(2021){Blagorodnova}, {Klencki}, {Pejcha},
  {Vreeswijk}, {Bond}, {Burdge}, {De}, {Fremling}, {Gehrz}, {Jencson},
  {Kasliwal}, {Kupfer}, {Lau}, {Masci}, \& {Rich}}]{2021A&A...653A.134B}
{Blagorodnova}, N., {Klencki}, J., {Pejcha}, O., {et~al.} 2021, \aap, 653,
  A134, \dodoi{10.1051/0004-6361/202140525}

\bibitem[{{Bond} {et~al.}(2003){Bond}, {Henden}, {Levay}, {Panagia}, {Sparks},
  {Starrfield}, {Wagner}, {Corradi}, \& {Munari}}]{2003Natur.422..405B}
{Bond}, H.~E., {Henden}, A., {Levay}, Z.~G., {et~al.} 2003, \nat, 422, 405,
  \dodoi{10.1038/nature01508}

\bibitem[{{Cai} {et~al.}(2019){Cai}, {Pastorello}, {Fraser}, {Prentice},
  {Reynolds}, {Cappellaro}, {Benetti}, {Morales-Garoffolo}, {Reguitti},
  {Elias-Rosa}, {Brennan}, {Callis}, {Cannizzaro}, {Fiore}, {Gromadzki},
  {Galindo-Guil}, {Gall}, {Heikkil{\"a}}, {Mason}, {Moran}, {Onori},
  {Sagu{\'e}s Carracedo}, \& {Valerin}}]{2019A&A...632L...6C}
{Cai}, Y.~Z., {Pastorello}, A., {Fraser}, M., {et~al.} 2019, \aap, 632, L6,
  \dodoi{10.1051/0004-6361/201936749}

\bibitem[{{Chesneau} {et~al.}(2014){Chesneau}, {Millour}, {De Marco}, {Bright},
  {Spang}, {Banerjee}, {Ashok}, {Kami{\'n}ski}, {Wisniewski}, {Meilland}, \&
  {Lagadec}}]{2014A&A...569L...3C}
{Chesneau}, O., {Millour}, F., {De Marco}, O., {et~al.} 2014, \aap, 569, L3,
  \dodoi{10.1051/0004-6361/201424458}

\bibitem[{{Choi} {et~al.}(2016){Choi}, {Dotter}, {Conroy}, {Cantiello},
  {Paxton}, \& {Johnson}}]{2016ApJ...823..102C}
{Choi}, J., {Dotter}, A., {Conroy}, C., {et~al.} 2016, \apj, 823, 102,
  \dodoi{10.3847/0004-637X/823/2/102}

\bibitem[{{De} {et~al.}(2020){De}, {Hankins}, {Kasliwal}, {Moore}, {Ofek},
  {Adams}, {Ashley}, {Babul}, {Bagdasaryan}, {Burdge}, {Burnham}, {Dekany},
  {Declacroix}, {Galla}, {Greffe}, {Hale}, {Jencson}, {Lau}, {Mahabal},
  {McKenna}, {Sharma}, {Shopbell}, {Smith}, {Soon}, {Sokoloski}, {Soria}, \&
  {Travouillon}}]{2020PASP..132b5001D}
{De}, K., {Hankins}, M.~J., {Kasliwal}, M.~M., {et~al.} 2020, \pasp, 132,
  025001, \dodoi{10.1088/1538-3873/ab6069}

\bibitem[{{De Marco} \& {Izzard}(2017)}]{2017PASA...34....1D}
{De Marco}, O., \& {Izzard}, R.~G. 2017, \pasa, 34, e001,
  \dodoi{10.1017/pasa.2016.52}

\bibitem[{{De Marco} {et~al.}(2011){De Marco}, {Passy}, {Moe}, {Herwig}, {Mac
  Low}, \& {Paxton}}]{2011MNRAS.411.2277D}
{De Marco}, O., {Passy}, J.-C., {Moe}, M., {et~al.} 2011, \mnras, 411, 2277,
  \dodoi{10.1111/j.1365-2966.2010.17891.x}

\bibitem[{{de Mink} {et~al.}(2013){de Mink}, {Langer}, {Izzard}, {Sana}, \& {de
  Koter}}]{2013ApJ...764..166D}
{de Mink}, S.~E., {Langer}, N., {Izzard}, R.~G., {Sana}, H., \& {de Koter}, A.
  2013, \apj, 764, 166, \dodoi{10.1088/0004-637X/764/2/166}

\bibitem[{{de Mink} {et~al.}(2014){de Mink}, {Sana}, {Langer}, {Izzard}, \&
  {Schneider}}]{2014ApJ...782....7D}
{de Mink}, S.~E., {Sana}, H., {Langer}, N., {Izzard}, R.~G., \& {Schneider},
  F.~R.~N. 2014, \apj, 782, 7, \dodoi{10.1088/0004-637X/782/1/7}

\bibitem[{{Dominik} {et~al.}(2012){Dominik}, {Belczynski}, {Fryer}, {Holz},
  {Berti}, {Bulik}, {Mandel}, \& {O'Shaughnessy}}]{2012ApJ...759...52D}
{Dominik}, M., {Belczynski}, K., {Fryer}, C., {et~al.} 2012, \apj, 759, 52,
  \dodoi{10.1088/0004-637X/759/1/52}

\bibitem[{{Duch{\^e}ne} \& {Kraus}(2013)}]{2013ARA&A..51..269D}
{Duch{\^e}ne}, G., \& {Kraus}, A. 2013, \araa, 51, 269,
  \dodoi{10.1146/annurev-astro-081710-102602}

\bibitem[{{Eggleton}(1983)}]{1983ApJ...268..368E}
{Eggleton}, P.~P. 1983, \apj, 268, 368, \dodoi{10.1086/160960}

\bibitem[{{Ferreira} {et~al.}(2019){Ferreira}, {Saito}, {Minniti}, {Navarro},
  {Ramos}, {Smith}, \& {Lucas}}]{2019MNRAS.486.1220F}
{Ferreira}, T., {Saito}, R.~K., {Minniti}, D., {et~al.} 2019, \mnras, 486,
  1220, \dodoi{10.1093/mnras/stz878}

\bibitem[{{Glanz} \& {Perets}(2018)}]{2018MNRAS.478L..12G}
{Glanz}, H., \& {Perets}, H.~B. 2018, \mnras, 478, L12,
  \dodoi{10.1093/mnrasl/sly065}

\bibitem[{{Hjellming} \& {Webbink}(1987)}]{1987ApJ...318..794H}
{Hjellming}, M.~S., \& {Webbink}, R.~F. 1987, \apj, 318, 794,
  \dodoi{10.1086/165412}

\bibitem[{{H{\"o}fner} {et~al.}(2016){H{\"o}fner}, {Bladh}, {Aringer}, \&
  {Ahuja}}]{2016A&A...594A.108H}
{H{\"o}fner}, S., {Bladh}, S., {Aringer}, B., \& {Ahuja}, R. 2016, \aap, 594,
  A108, \dodoi{10.1051/0004-6361/201628424}

\bibitem[{{H{\"o}fner} {et~al.}(2022){H{\"o}fner}, {Bladh}, {Aringer}, \&
  {Eriksson}}]{2022A&A...657A.109H}
{H{\"o}fner}, S., {Bladh}, S., {Aringer}, B., \& {Eriksson}, K. 2022, \aap,
  657, A109, \dodoi{10.1051/0004-6361/202141224}

\bibitem[{{Hubov{\'a}} \& {Pejcha}(2019)}]{2019MNRAS.489..891H}
{Hubov{\'a}}, D., \& {Pejcha}, O. 2019, \mnras, 489, 891,
  \dodoi{10.1093/mnras/stz2208}

\bibitem[{Hunter(2007)}]{Hunter:2007}
Hunter, J.~D. 2007, Computing In Science \& Engineering, 9, 90

\bibitem[{{Iaconi} {et~al.}(2019){Iaconi}, {Maeda}, {De Marco}, {Nozawa}, \&
  {Reichardt}}]{2019MNRAS.489.3334I}
{Iaconi}, R., {Maeda}, K., {De Marco}, O., {Nozawa}, T., \& {Reichardt}, T.
  2019, \mnras, 489, 3334, \dodoi{10.1093/mnras/stz2312}

\bibitem[{{Iaconi} {et~al.}(2020){Iaconi}, {Maeda}, {Nozawa}, {De Marco}, \&
  {Reichardt}}]{2020MNRAS.497.3166I}
{Iaconi}, R., {Maeda}, K., {Nozawa}, T., {De Marco}, O., \& {Reichardt}, T.
  2020, \mnras, 497, 3166, \dodoi{10.1093/mnras/staa2169}

\bibitem[{{Iben} \& {Livio}(1993)}]{1993PASP..105.1373I}
{Iben}, Icko, J., \& {Livio}, M. 1993, \pasp, 105, 1373, \dodoi{10.1086/133321}

\bibitem[{{Ivanova} {et~al.}(2013){Ivanova}, {Justham}, {Chen}, {De Marco},
  {Fryer}, {Gaburov}, {Ge}, {Glebbeek}, {Han}, {Li}, {Lu}, {Marsh},
  {Podsiadlowski}, {Potter}, {Soker}, {Taam}, {Tauris}, {van den Heuvel}, \&
  {Webbink}}]{2013A&ARv..21...59I}
{Ivanova}, N., {Justham}, S., {Chen}, X., {et~al.} 2013, \aapr, 21, 59,
  \dodoi{10.1007/s00159-013-0059-2}

\bibitem[{{Jencson} {et~al.}(2019){Jencson}, {Kasliwal}, {Adams}, {Bond}, {De},
  {Johansson}, {Karambelkar}, {Lau}, {Tinyanont}, {Ryder}, {Cody}, {Masci},
  {Bally}, {Blagorodnova}, {Castell{\'o}n}, {Fremling}, {Gehrz}, {Helou},
  {Kilpatrick}, {Milne}, {Morrell}, {Perley}, {Phillips}, {Smith}, {van Dyk},
  \& {Williams}}]{2019ApJ...886...40J}
{Jencson}, J.~E., {Kasliwal}, M.~M., {Adams}, S.~M., {et~al.} 2019, \apj, 886,
  40, \dodoi{10.3847/1538-4357/ab4a01}

\bibitem[{{Kami{\'n}ski}(2008)}]{2008A&A...482..803K}
{Kami{\'n}ski}, T. 2008, \aap, 482, 803, \dodoi{10.1051/0004-6361:20079189}

\bibitem[{{Kami{\'n}ski} {et~al.}(2015){Kami{\'n}ski}, {Mason}, {Tylenda}, \&
  {Schmidt}}]{2015A&A...580A..34K}
{Kami{\'n}ski}, T., {Mason}, E., {Tylenda}, R., \& {Schmidt}, M.~R. 2015, \aap,
  580, A34, \dodoi{10.1051/0004-6361/201526212}

\bibitem[{{Kami{\'n}ski} {et~al.}(2007){Kami{\'n}ski}, {Miller}, \&
  {Tylenda}}]{2007A&A...475..569K}
{Kami{\'n}ski}, T., {Miller}, M., \& {Tylenda}, R. 2007, \aap, 475, 569,
  \dodoi{10.1051/0004-6361:20077982}

\bibitem[{{Kami{\'n}ski} {et~al.}(2010){Kami{\'n}ski}, {Schmidt}, \&
  {Tylenda}}]{2010A&A...522A..75K}
{Kami{\'n}ski}, T., {Schmidt}, M., \& {Tylenda}, R. 2010, \aap, 522, A75,
  \dodoi{10.1051/0004-6361/201014406}

\bibitem[{{Kami{\'n}ski} {et~al.}(2009){Kami{\'n}ski}, {Schmidt}, {Tylenda},
  {Konacki}, \& {Gromadzki}}]{2009ApJS..182...33K}
{Kami{\'n}ski}, T., {Schmidt}, M., {Tylenda}, R., {Konacki}, M., \&
  {Gromadzki}, M. 2009, \apjs, 182, 33, \dodoi{10.1088/0067-0049/182/1/33}

\bibitem[{{Kami{\'n}ski} {et~al.}(2018){Kami{\'n}ski}, {Steffen}, {Tylenda},
  {Young}, {Patel}, \& {Menten}}]{2018A&A...617A.129K}
{Kami{\'n}ski}, T., {Steffen}, W., {Tylenda}, R., {et~al.} 2018, \aap, 617,
  A129, \dodoi{10.1051/0004-6361/201833165}

\bibitem[{{Kami{\'n}ski} \& {Tylenda}(2011)}]{2011A&A...527A..75K}
{Kami{\'n}ski}, T., \& {Tylenda}, R. 2011, \aap, 527, A75,
  \dodoi{10.1051/0004-6361/201015950}

\bibitem[{{Kami{\'n}ski} \& {Tylenda}(2013)}]{2013A&A...558A..82K}
---. 2013, \aap, 558, A82, \dodoi{10.1051/0004-6361/201321852}

\bibitem[{{Kami{\'n}ski} {et~al.}(2021){Kami{\'n}ski}, {Tylenda}, {Kiljan},
  {Schmidt}, {Lisiecki}, {Melis}, {Frankowski}, {Joshi}, \&
  {Menten}}]{2021A&A...655A..32K}
{Kami{\'n}ski}, T., {Tylenda}, R., {Kiljan}, A., {et~al.} 2021, \aap, 655, A32,
  \dodoi{10.1051/0004-6361/202141526}

\bibitem[{{Kasliwal} {et~al.}(2017){Kasliwal}, {Bally}, {Masci}, {Cody},
  {Bond}, {Jencson}, {Tinyanont}, {Cao}, {Contreras}, {Dykhoff}, {Amodeo},
  {Armus}, {Boyer}, {Cantiello}, {Carlon}, {Cass}, {Cook}, {Corgan}, {Faella},
  {Fox}, {Green}, {Gehrz}, {Helou}, {Hsiao}, {Johansson}, {Khan}, {Lau},
  {Langer}, {Levesque}, {Milne}, {Mohamed}, {Morrell}, {Monson}, {Moore},
  {Ofek}, {O' Sullivan}, {Parthasarathy}, {Perez}, {Perley}, {Phillips},
  {Prince}, {Shenoy}, {Smith}, {Surace}, {Van Dyk}, {Whitelock}, \&
  {Williams}}]{2017ApJ...839...88K}
{Kasliwal}, M.~M., {Bally}, J., {Masci}, F., {et~al.} 2017, \apj, 839, 88,
  \dodoi{10.3847/1538-4357/aa6978}

\bibitem[{{Kimeswenger} {et~al.}(2002){Kimeswenger}, {Lederle}, {Schmeja}, \&
  {Armsdorfer}}]{2002MNRAS.336L..43K}
{Kimeswenger}, S., {Lederle}, C., {Schmeja}, S., \& {Armsdorfer}, B. 2002,
  \mnras, 336, L43, \dodoi{10.1046/j.1365-8711.2002.06017.x}

\bibitem[{{Kochanek} {et~al.}(2014){Kochanek}, {Adams}, \&
  {Belczynski}}]{2014MNRAS.443.1319K}
{Kochanek}, C.~S., {Adams}, S.~M., \& {Belczynski}, K. 2014, \mnras, 443, 1319,
  \dodoi{10.1093/mnras/stu1226}

\bibitem[{{Kurtenkov} {et~al.}(2015){Kurtenkov}, {Pessev}, {Tomov},
  {Barsukova}, {Fabrika}, {Vida}, {Hornoch}, {Ovcharov}, {Goranskij}, {Valeev},
  {Moln{\'a}r}, {S{\'a}rneczky}, {Kostov}, {Nedialkov}, {Valenti}, {Geier},
  {Wiersema}, {Henze}, {Shafter}, {Mu{\~n}oz Dimitrova}, {Popov}, \&
  {Stritzinger}}]{2015A&A...578L..10K}
{Kurtenkov}, A.~A., {Pessev}, P., {Tomov}, T., {et~al.} 2015, \aap, 578, L10,
  \dodoi{10.1051/0004-6361/201526564}

\bibitem[{{L{\"u}} {et~al.}(2013){L{\"u}}, {Zhu}, \&
  {Podsiadlowski}}]{2013ApJ...768..193L}
{L{\"u}}, G., {Zhu}, C., \& {Podsiadlowski}, P. 2013, \apj, 768, 193,
  \dodoi{10.1088/0004-637X/768/2/193}

\bibitem[{MacLeod(2020)}]{RLOF1.1}
MacLeod, M. 2020, morganemacleod/RLOF: v1.1, v.1.1,  Zenodo,
  \dodoi{10.5281/zenodo.3690127}

\bibitem[{MacLeod(2022)}]{morgan_macleod_2022_6554801}
---. 2022, morganemacleod/dustytransients\_materials: v0.1, v0.1,  Zenodo,
  \dodoi{10.5281/zenodo.6554801}

\bibitem[{{MacLeod} \& {Loeb}(2020{\natexlab{a}})}]{2020ApJ...893..106M}
{MacLeod}, M., \& {Loeb}, A. 2020{\natexlab{a}}, \apj, 893, 106,
  \dodoi{10.3847/1538-4357/ab822e}

\bibitem[{{MacLeod} \& {Loeb}(2020{\natexlab{b}})}]{2020ApJ...895...29M}
---. 2020{\natexlab{b}}, \apj, 895, 29, \dodoi{10.3847/1538-4357/ab89b6}

\bibitem[{{MacLeod} {et~al.}(2017){MacLeod}, {Macias}, {Ramirez-Ruiz},
  {Grindlay}, {Batta}, \& {Montes}}]{2017ApJ...835..282M}
{MacLeod}, M., {Macias}, P., {Ramirez-Ruiz}, E., {et~al.} 2017, \apj, 835, 282,
  \dodoi{10.3847/1538-4357/835/2/282}

\bibitem[{{MacLeod} {et~al.}(2018{\natexlab{a}}){MacLeod}, {Ostriker}, \&
  {Stone}}]{2018ApJ...863....5M}
{MacLeod}, M., {Ostriker}, E.~C., \& {Stone}, J.~M. 2018{\natexlab{a}}, \apj,
  863, 5, \dodoi{10.3847/1538-4357/aacf08}

\bibitem[{{MacLeod} {et~al.}(2018{\natexlab{b}}){MacLeod}, {Ostriker}, \&
  {Stone}}]{2018ApJ...868..136M}
---. 2018{\natexlab{b}}, \apj, 868, 136, \dodoi{10.3847/1538-4357/aae9eb}

\bibitem[{{Marchant} {et~al.}(2021){Marchant}, {Pappas}, {Gallegos-Garcia},
  {Berry}, {Taam}, {Kalogera}, \& {Podsiadlowski}}]{2021A&A...650A.107M}
{Marchant}, P., {Pappas}, K. M.~W., {Gallegos-Garcia}, M., {et~al.} 2021, \aap,
  650, A107, \dodoi{10.1051/0004-6361/202039992}

\bibitem[{{Martini} {et~al.}(1999){Martini}, {Wagner}, {Tomaney}, {Rich},
  {della Valle}, \& {Hauschildt}}]{1999AJ....118.1034M}
{Martini}, P., {Wagner}, R.~M., {Tomaney}, A., {et~al.} 1999, \aj, 118, 1034,
  \dodoi{10.1086/300951}

\bibitem[{{Mason} {et~al.}(2010){Mason}, {Diaz}, {Williams}, {Preston}, \&
  {Bensby}}]{2010A&A...516A.108M}
{Mason}, E., {Diaz}, M., {Williams}, R.~E., {Preston}, G., \& {Bensby}, T.
  2010, \aap, 516, A108, \dodoi{10.1051/0004-6361/200913610}

\bibitem[{{Matsumoto} \& {Metzger}(2022)}]{2022arXiv220210478M}
{Matsumoto}, T., \& {Metzger}, B.~D. 2022, arXiv e-prints, arXiv:2202.10478.
\newblock \doarXiv{2202.10478}

\bibitem[{{Mauerhan} {et~al.}(2015){Mauerhan}, {Van Dyk}, {Graham}, {Zheng},
  {Clubb}, {Filippenko}, {Valenti}, {Brown}, {Smith}, {Howell}, \&
  {Arcavi}}]{2015MNRAS.447.1922M}
{Mauerhan}, J.~C., {Van Dyk}, S.~D., {Graham}, M.~L., {et~al.} 2015, \mnras,
  447, 1922, \dodoi{10.1093/mnras/stu2541}

\bibitem[{{McCollum} {et~al.}(2014){McCollum}, {Laine}, {V{\"a}is{\"a}nen},
  {Bruhweiler}, {Rottler}, {Ryder}, {Wahlgren}, {Barway}, {Nagayama}, \&
  {Ramphul}}]{2014AJ....147...11M}
{McCollum}, B., {Laine}, S., {V{\"a}is{\"a}nen}, P., {et~al.} 2014, \aj, 147,
  11, \dodoi{10.1088/0004-6256/147/1/11}

\bibitem[{{Melis}(2020)}]{2020RNAAS...4..238M}
{Melis}, C. 2020, Research Notes of the American Astronomical Society, 4, 238,
  \dodoi{10.3847/2515-5172/abd32a}

\bibitem[{{Metzger} \& {Pejcha}(2017)}]{2017MNRAS.471.3200M}
{Metzger}, B.~D., \& {Pejcha}, O. 2017, \mnras, 471, 3200,
  \dodoi{10.1093/mnras/stx1768}

\bibitem[{{Moe} \& {Di Stefano}(2017)}]{2017ApJS..230...15M}
{Moe}, M., \& {Di Stefano}, R. 2017, \apjs, 230, 15,
  \dodoi{10.3847/1538-4365/aa6fb6}

\bibitem[{{Munari} {et~al.}(2002){Munari}, {Henden}, {Kiyota}, {Laney},
  {Marang}, {Zwitter}, {Corradi}, {Desidera}, {Marrese}, {Giro}, {Boschi}, \&
  {Schwartz}}]{2002A&A...389L..51M}
{Munari}, U., {Henden}, A., {Kiyota}, S., {et~al.} 2002, \aap, 389, L51,
  \dodoi{10.1051/0004-6361:20020715}

\bibitem[{{Nicholls} {et~al.}(2013){Nicholls}, {Melis}, {Soszynski}, {Udalski},
  {Szymanski}, {Kubiak}, {Pietrzynski}, {Poleski}, {Ulaczyk}, {Wyrzykowski},
  {Kozlowski}, \& {Pietrukowicz}}]{2013MNRAS.431L..33N}
{Nicholls}, C.~P., {Melis}, C., {Soszynski}, I., {et~al.} 2013, \mnras, 431,
  L33, \dodoi{10.1093/mnrasl/slt003}

\bibitem[{{{\"O}pik}(1924)}]{1924PTarO..25f...1O}
{{\"O}pik}, E. 1924, Publications of the Tartu Astrofizica Observatory, 25, 1

\bibitem[{{Paczynski}(1976)}]{1976IAUS...73...75P}
{Paczynski}, B. 1976, in Structure and Evolution of Close Binary Systems, ed.
  P.~{Eggleton}, S.~{Mitton}, \& J.~{Whelan}, Vol.~73, 75

\bibitem[{{Paczy{\'n}ski} \& {Sienkiewicz}(1972)}]{1972AcA....22...73P}
{Paczy{\'n}ski}, B., \& {Sienkiewicz}, R. 1972, \actaa, 22, 73

\bibitem[{{Pastorello} {et~al.}(2021{\natexlab{a}}){Pastorello}, {Fraser},
  {Valerin}, {Reguitti}, {Itagaki}, {Ochner}, {Williams}, {Jones}, {Munday},
  {Smartt}, {Smith}, {Srivastav}, {Elias-Rosa}, {Kankare}, {Karamehmetoglu},
  {Lundqvist}, {Mazzali}, {Munari}, {Stritzinger}, {Tomasella}, {Anderson},
  {Chambers}, \& {Rest}}]{2021A&A...646A.119P}
{Pastorello}, A., {Fraser}, M., {Valerin}, G., {et~al.} 2021{\natexlab{a}},
  \aap, 646, A119, \dodoi{10.1051/0004-6361/202039952}

\bibitem[{{Pastorello} {et~al.}(2021{\natexlab{b}}){Pastorello}, {Valerin},
  {Fraser}, {Elias-Rosa}, {Valenti}, {Reguitti}, {Mazzali}, {Amaro}, {Andrews},
  {Dong}, {Jencson}, {Lundquist}, {Reichart}, {Sand}, {Wyatt}, {Smartt},
  {Smith}, {Srivastav}, {Cai}, {Cappellaro}, {Holmbo}, {Fiore}, {Jones},
  {Kankare}, {Karamehmetoglu}, {Lundqvist}, {Morales-Garoffolo}, {Reynolds},
  {Stritzinger}, {Williams}, {Chambers}, {de Boer}, {Huber}, {Rest}, \&
  {Wainscoat}}]{2021A&A...647A..93P}
{Pastorello}, A., {Valerin}, G., {Fraser}, M., {et~al.} 2021{\natexlab{b}},
  \aap, 647, A93, \dodoi{10.1051/0004-6361/202039953}

\bibitem[{{Paxton} {et~al.}(2011){Paxton}, {Bildsten}, {Dotter}, {Herwig},
  {Lesaffre}, \& {Timmes}}]{2011ApJS..192....3P}
{Paxton}, B., {Bildsten}, L., {Dotter}, A., {et~al.} 2011, \apjs, 192, 3,
  \dodoi{10.1088/0067-0049/192/1/3}

\bibitem[{{Paxton} {et~al.}(2013){Paxton}, {Cantiello}, {Arras}, {Bildsten},
  {Brown}, {Dotter}, {Mankovich}, {Montgomery}, {Stello}, {Timmes}, \&
  {Townsend}}]{2013ApJS..208....4P}
{Paxton}, B., {Cantiello}, M., {Arras}, P., {et~al.} 2013, \apjs, 208, 4,
  \dodoi{10.1088/0067-0049/208/1/4}

\bibitem[{{Paxton} {et~al.}(2015){Paxton}, {Marchant}, {Schwab}, {Bauer},
  {Bildsten}, {Cantiello}, {Dessart}, {Farmer}, {Hu}, {Langer}, {Townsend},
  {Townsley}, \& {Timmes}}]{2015ApJS..220...15P}
{Paxton}, B., {Marchant}, P., {Schwab}, J., {et~al.} 2015, \apjs, 220, 15,
  \dodoi{10.1088/0067-0049/220/1/15}

\bibitem[{{Paxton} {et~al.}(2018){Paxton}, {Schwab}, {Bauer}, {Bildsten},
  {Blinnikov}, {Duffell}, {Farmer}, {Goldberg}, {Marchant}, {Sorokina},
  {Thoul}, {Townsend}, \& {Timmes}}]{2018ApJS..234...34P}
{Paxton}, B., {Schwab}, J., {Bauer}, E.~B., {et~al.} 2018, \apjs, 234, 34,
  \dodoi{10.3847/1538-4365/aaa5a8}

\bibitem[{{Paxton} {et~al.}(2019){Paxton}, {Smolec}, {Schwab}, {Gautschy},
  {Bildsten}, {Cantiello}, {Dotter}, {Farmer}, {Goldberg}, {Jermyn}, {Kanbur},
  {Marchant}, {Thoul}, {Townsend}, {Wolf}, {Zhang}, \&
  {Timmes}}]{2019ApJS..243...10P}
{Paxton}, B., {Smolec}, R., {Schwab}, J., {et~al.} 2019, \apjs, 243, 10,
  \dodoi{10.3847/1538-4365/ab2241}

\bibitem[{{Pejcha}(2014)}]{2014ApJ...788...22P}
{Pejcha}, O. 2014, \apj, 788, 22, \dodoi{10.1088/0004-637X/788/1/22}

\bibitem[{{Pejcha} {et~al.}(2016{\natexlab{a}}){Pejcha}, {Metzger}, \&
  {Tomida}}]{2016MNRAS.455.4351P}
{Pejcha}, O., {Metzger}, B.~D., \& {Tomida}, K. 2016{\natexlab{a}}, \mnras,
  455, 4351, \dodoi{10.1093/mnras/stv2592}

\bibitem[{{Pejcha} {et~al.}(2016{\natexlab{b}}){Pejcha}, {Metzger}, \&
  {Tomida}}]{2016MNRAS.461.2527P}
---. 2016{\natexlab{b}}, \mnras, 461, 2527, \dodoi{10.1093/mnras/stw1481}

\bibitem[{{Pejcha} {et~al.}(2017){Pejcha}, {Metzger}, {Tyles}, \&
  {Tomida}}]{2017ApJ...850...59P}
{Pejcha}, O., {Metzger}, B.~D., {Tyles}, J.~G., \& {Tomida}, K. 2017, \apj,
  850, 59, \dodoi{10.3847/1538-4357/aa95b9}

\bibitem[{P\'erez \& Granger(2007)}]{PER-GRA:2007}
P\'erez, F., \& Granger, B.~E. 2007, Computing in Science and Engineering, 9,
  21, \dodoi{10.1109/MCSE.2007.53}

\bibitem[{{Postnov} \& {Yungelson}(2014)}]{2014LRR....17....3P}
{Postnov}, K.~A., \& {Yungelson}, L.~R. 2014, Living Reviews in Relativity, 17,
  3, \dodoi{10.12942/lrr-2014-3}

\bibitem[{{Sana} {et~al.}(2012){Sana}, {de Mink}, {de Koter}, {Langer},
  {Evans}, {Gieles}, {Gosset}, {Izzard}, {Le Bouquin}, \&
  {Schneider}}]{2012Sci...337..444S}
{Sana}, H., {de Mink}, S.~E., {de Koter}, A., {et~al.} 2012, Science, 337, 444,
  \dodoi{10.1126/science.1223344}

\bibitem[{{Schneider} {et~al.}(2016){Schneider}, {Podsiadlowski}, {Langer},
  {Castro}, \& {Fossati}}]{2016MNRAS.457.2355S}
{Schneider}, F.~R.~N., {Podsiadlowski}, P., {Langer}, N., {Castro}, N., \&
  {Fossati}, L. 2016, \mnras, 457, 2355, \dodoi{10.1093/mnras/stw148}

\bibitem[{{Shara} {et~al.}(1985){Shara}, {Moffat}, \&
  {Webbink}}]{1985ApJ...294..271S}
{Shara}, M.~M., {Moffat}, A.~F.~J., \& {Webbink}, R.~F. 1985, \apj, 294, 271,
  \dodoi{10.1086/163296}

\bibitem[{{Smith} {et~al.}(2016){Smith}, {Andrews}, {Van Dyk}, {Mauerhan},
  {Kasliwal}, {Bond}, {Filippenko}, {Clubb}, {Graham}, {Perley}, {Jencson},
  {Bally}, {Ubeda}, \& {Sabbi}}]{2016MNRAS.458..950S}
{Smith}, N., {Andrews}, J.~E., {Van Dyk}, S.~D., {et~al.} 2016, \mnras, 458,
  950, \dodoi{10.1093/mnras/stw219}

\bibitem[{{Soker} \& {Tylenda}(2003)}]{2003ApJ...582L.105S}
{Soker}, N., \& {Tylenda}, R. 2003, \apjl, 582, L105, \dodoi{10.1086/367759}

\bibitem[{{Soker} \& {Tylenda}(2006)}]{2006MNRAS.373..733S}
---. 2006, \mnras, 373, 733, \dodoi{10.1111/j.1365-2966.2006.11056.x}

\bibitem[{{Soker} \& {Tylenda}(2007)}]{2007MNRAS.375..909S}
---. 2007, \mnras, 375, 909, \dodoi{10.1111/j.1365-2966.2006.11351.x}

\bibitem[{{Tylenda}(2005)}]{2005A&A...436.1009T}
{Tylenda}, R. 2005, \aap, 436, 1009, \dodoi{10.1051/0004-6361:20052800}

\bibitem[{{Tylenda} {et~al.}(2005{\natexlab{a}}){Tylenda}, {Crause},
  {G{\'o}rny}, \& {Schmidt}}]{2005A&A...439..651T}
{Tylenda}, R., {Crause}, L.~A., {G{\'o}rny}, S.~K., \& {Schmidt}, M.~R.
  2005{\natexlab{a}}, \aap, 439, 651, \dodoi{10.1051/0004-6361:20041581}

\bibitem[{{Tylenda} {et~al.}(2015){Tylenda}, {G{\'o}rny}, {Kami{\'n}ski}, \&
  {Schmidt}}]{2015A&A...578A..75T}
{Tylenda}, R., {G{\'o}rny}, S.~K., {Kami{\'n}ski}, T., \& {Schmidt}, M. 2015,
  \aap, 578, A75, \dodoi{10.1051/0004-6361/201425592}

\bibitem[{{Tylenda} \& {Kami{\'n}ski}(2016)}]{2016A&A...592A.134T}
{Tylenda}, R., \& {Kami{\'n}ski}, T. 2016, \aap, 592, A134,
  \dodoi{10.1051/0004-6361/201527700}

\bibitem[{{Tylenda} \& {Soker}(2006)}]{2006A&A...451..223T}
{Tylenda}, R., \& {Soker}, N. 2006, \aap, 451, 223,
  \dodoi{10.1051/0004-6361:20054201}

\bibitem[{{Tylenda} {et~al.}(2005{\natexlab{b}}){Tylenda}, {Soker}, \&
  {Szczerba}}]{2005A&A...441.1099T}
{Tylenda}, R., {Soker}, N., \& {Szczerba}, R. 2005{\natexlab{b}}, \aap, 441,
  1099, \dodoi{10.1051/0004-6361:20042485}

\bibitem[{{Tylenda} {et~al.}(2011){Tylenda}, {Hajduk}, {Kami{\'n}ski},
  {Udalski}, {Soszy{\'n}ski}, {Szyma{\'n}ski}, {Kubiak}, {Pietrzy{\'n}ski},
  {Poleski}, {Wyrzykowski}, \& {Ulaczyk}}]{2011A&A...528A.114T}
{Tylenda}, R., {Hajduk}, M., {Kami{\'n}ski}, T., {et~al.} 2011, \aap, 528,
  A114, \dodoi{10.1051/0004-6361/201016221}

\bibitem[{{Tylenda} {et~al.}(2013){Tylenda}, {Kami{\'n}ski}, {Udalski},
  {Soszy{\'n}ski}, {Poleski}, {Szyma{\'n}ski}, {Kubiak}, {Pietrzy{\'n}ski},
  {Koz{\l}owski}, {Pietrukowicz}, {Ulaczyk}, \&
  {Wyrzykowski}}]{2013A&A...555A..16T}
{Tylenda}, R., {Kami{\'n}ski}, T., {Udalski}, A., {et~al.} 2013, \aap, 555,
  A16, \dodoi{10.1051/0004-6361/201321647}

\bibitem[{{van den Heuvel}(1976)}]{1976IAUS...73...35V}
{van den Heuvel}, E.~P.~J. 1976, in Structure and Evolution of Close Binary
  Systems, ed. P.~{Eggleton}, S.~{Mitton}, \& J.~{Whelan}, Vol.~73, 35

\bibitem[{Van Der~Walt {et~al.}(2011)Van Der~Walt, Colbert, \&
  Varoquaux}]{van2011numpy}
Van Der~Walt, S., Colbert, S.~C., \& Varoquaux, G. 2011, Computing in Science
  \& Engineering, 13, 22

\bibitem[{{Vigna-G{\'o}mez} {et~al.}(2020){Vigna-G{\'o}mez}, {MacLeod},
  {Neijssel}, {Broekgaarden}, {Justham}, {Howitt}, {de Mink}, {Vinciguerra}, \&
  {Mandel}}]{2020PASA...37...38V}
{Vigna-G{\'o}mez}, A., {MacLeod}, M., {Neijssel}, C.~J., {et~al.} 2020, \pasa,
  37, e038, \dodoi{10.1017/pasa.2020.31}

\bibitem[{{Virtanen} {et~al.}(2020){Virtanen}, {Gommers}, {Oliphant},
  {Haberland}, {Reddy}, {Cournapeau}, {Burovski}, {Peterson}, {Weckesser},
  {Bright}, {van der Walt}, {Brett}, {Wilson}, {Jarrod Millman}, {Mayorov},
  {Nelson}, {Jones}, {Kern}, {Larson}, {Carey}, {Polat}, {Feng}, {Moore}, {Vand
  erPlas}, {Laxalde}, {Perktold}, {Cimrman}, {Henriksen}, {Quintero}, {Harris},
  {Archibald}, {Ribeiro}, {Pedregosa}, {van Mulbregt}, \&
  {Contributors}}]{2020SciPy-NMeth}
{Virtanen}, P., {Gommers}, R., {Oliphant}, T.~E., {et~al.} 2020, Nature
  Methods, \dodoi{https://doi.org/10.1038/s41592-019-0686-2}

\bibitem[{{Wang} {et~al.}(2015){Wang}, {Zhu}, \&
  {L{\"u}}}]{2015RAA....15...55W}
{Wang}, Z.-J., {Zhu}, C.-H., \& {L{\"u}}, G.-L. 2015, Research in Astronomy and
  Astrophysics, 15, 55, \dodoi{10.1088/1674-4527/15/1/006}

\bibitem[{{Webbink}(1984)}]{1984ApJ...277..355W}
{Webbink}, R.~F. 1984, \apj, 277, 355, \dodoi{10.1086/161701}

\bibitem[{{Williams} {et~al.}(2015){Williams}, {Darnley}, {Bode}, \&
  {Steele}}]{2015ApJ...805L..18W}
{Williams}, S.~C., {Darnley}, M.~J., {Bode}, M.~F., \& {Steele}, I.~A. 2015,
  \apjl, 805, L18, \dodoi{10.1088/2041-8205/805/2/L18}

\bibitem[{{Woodward} {et~al.}(2021){Woodward}, {Evans}, {Banerjee}, {Liimets},
  {Djupvik}, {Starrfield}, {Clayton}, {Eyres}, {Gehrz}, \&
  {Wagner}}]{2021AJ....162..183W}
{Woodward}, C.~E., {Evans}, A., {Banerjee}, D.~P.~K., {et~al.} 2021, \aj, 162,
  183, \dodoi{10.3847/1538-3881/ac1f1e}

\bibitem[{{Zhu} {et~al.}(2013){Zhu}, {L{\"u}}, \& {Wang}}]{2013ApJ...777...23Z}
{Zhu}, C., {L{\"u}}, G., \& {Wang}, Z. 2013, \apj, 777, 23,
  \dodoi{10.1088/0004-637X/777/1/23}

\end{thebibliography}


\end{document}